\documentclass[preprint2,tighten]{aastex62_reep}

\usepackage{epstopdf}
\usepackage{subfigure}  
\usepackage{amsmath}
\usepackage{hyperref}
\usepackage{xcolor}
\usepackage{soul}
\usepackage{dutchcal}
\usepackage[normalem]{ulem}
\shorttitle{Low-FIP Injection}
\shortauthors{Reep \& To}

\begin{document}

\title{Direct Injection of Chemically-Enhanced Plasma into a Solar Flare Arcade}

\author[0000-0003-4739-1152]{Jeffrey W. Reep}
\affiliation{Institute for Astronomy, University of Hawai'i at M\=anoa, Pukalani, HI 96768, USA}
\email{reep@hawaii.edu}

\author[0000-0003-0774-9084]{Andy S. H. To}
\affiliation{Université Paris-Saclay, CNRS, Institut d’Astrophysique Spatiale Orsay, France}
\affiliation{European Space Agency (ESA), European Space Research and Technology Centre (ESTEC), Keplerlaan 1, 2201 AZ Noordwijk, the Netherlands}

\begin{abstract}
The magnetic reconnection event that drives ``standard'' solar flares ejects material from the reconnection site, loading mass directly onto the newly-reconnected flare arcade.  \citet{to2024} suggested that this model can explain observations of the distribution of elemental abundances measured in post-flare loops: the reconnection exhaust directly injects plasma with coronal abundances onto the arcade, flowing from the apex downwards.  We test this idea directly by modeling the spatiotemporal variation of elements with low first ionization potential (FIP), and directly inject material near loop apices.  We show that this direct injection is capable of producing enhancements similar in magnitude and spatial spread across the corona to those observed during the gradual phase of an X11.9 flare.  We also show that direct chromospheric heating strongly limits the width of the low-FIP enhancement to a narrow region near the injection site.  The enhancement of low-FIP elemental abundances naturally explains bright knots commonly observed near the top of flare arcades.  
\end{abstract}


\section{Introduction}
\label{sec:intro}

Solar flares are driven by magnetic reconnection, where the conversion of magnetic energy into kinetic and thermal energy is released directly into a newly reconnected arcade.  This scenario was suggested by \citet{carmichael1964}, who proposed that accelerated particles are directly injected into a flare arcade, forming the basis for a large portion of modern hydrodynamic modeling of flares.  The direct injection of material was considered by \citet{kopp1976}, who argue that ``a downward propagating shock is set up on [recently closed flux tubes] heating and compressing the gas behind.''  Such shocks have been observed in large flares, often in limb flares where the current sheet may be more readily apparent \citep{liu2013,reeves2015,zhu2016,yu2020}.  These termination shocks are turbulent, with temperatures exceeding 10 MK and velocities of a few hundred km s$^{-1}$ \citep{doschek2014,reeves2015,li2018,warren2018,polito2018,yu2020,french2024}. 

In addition to termination shocks, supra-arcade downflows (SADs) are regularly observed as dark blobs, often described as tad-pole shaped, descending onto the flare arcade \citep{mckenzie1999}, at speeds ranging from only a few km s$^{-1}$ to hundreds of km s$^{-1}$ \citep{sheeley2004,warren2011,xie2022}.  These have been interpreted as direct observations of newly reconnected loops in the arcade, with each SAD ``pinched off'' to form a new loop \citep{sheeley2004}.  The loops, it is argued, then brighten due to chromospheric evaporation, which carries hot and dense material into the corona.  \citet{savage2012} suggested instead that SADs were voids left in the wakes of contracting loops, while \citet{french2025} argue that they are true down-flowing structures.  \citet{french2025} additionally measured the temperatures of SADs to be $\approx$ 12.2 to 13.4 MK, and find that they are more turbulent than the surrounding structure.  \citet{xie2023} argue that SADs heat the surrounding plasma, and, being turbulent, suppress conductive cooling.  

Observations of the 10 September 2017 X11.9\footnote{This event is often listed as X8.2 based on scaled GOES-15 observations.  The true irradiance value is X11.9 in GOES-15 data or X12.9 in GOES-16.  The confusion stems from a scaling factor of 0.7 that NOAA had previously included in XRS-B data, removed in 2020.} flare showed both a termination shock \citep{cheng2018,yu2020} as well as SADs \citep{longcope2018}.  The contraction of newly reconnected loops, additionally, continued for hours past the peak of the event \citep{hayes2019}.  Observations of the event in microwave, hard X-rays, and the extreme ultraviolet consistently showed downflowing plasma onto the solar flare arcade.  

Curiously, \citet{doschek2018} showed that the top of the flare arcade in this event exhibited an enhanced \ion{Ca}{14}/\ion{Ar}{14} line ratio (peak ionization $\approx$ 3.5 MK), suggesting that Ca was locally more abundant.  Elemental abundances of elements with low first ionization potential (FIP; \textit{e.g.} Ca, Fe, Mg, Si) are commonly observed to vary in the solar atmosphere, including during solar flares.  In the quiescent corona and solar wind, these elements are often enhanced by a factor of 3 to 5 relative to their proportions in the photosphere, and this enhancement is independent of atomic mass, an observational problem called the FIP effect \citep{feldman1998,feldman2000,laming2015,brooks2026}.  A popular explanation is that the ponderomotive force from downward-propagating Alfv\'en waves preferentially accelerates elements readily ionized in the chromosphere upwards into the corona, while not affecting neutrals \citep{laming2004,laming2021}.  Recent work has suggested that turbulence in the chromosphere may regulate the FIP effect, additionally \citep{to2026}.

In solar flares, however, low-FIP abundances have shown widely varying measurements, sometimes found to be photospheric \citep{warren2014}, sometimes coronal \citep{phillips2012,dennis2015}, sometimes showing an inverse FIP effect where high-FIP elements are depleted \citep{fludra1995,doschek2015,katsuda2020}, or, recently, abundances showing significant time-variation and complexity \citep{mondal2021,to2021,suzuki2025}.   Building upon the work of \citet{doschek2018}, the study by \citet{to2024} thoroughly examined the spatial and temporal variation of elemental abundances in the 10 September 2017 flare.  They examined both the \ion{Ca}{14}/\ion{Ar}{14} and \ion{Fe}{16}/\ion{S}{13} line ratios, which form in slightly different temperature ranges, in two loop structures in the face-on arcade.  During the gradual phase, the higher loop showed a stronger enhancement than the lower loop with both ratios.  At both heights and in both line ratios, additionally, the abundance factor was peaked near the apex, \textit{gradually} falling to photospheric levels nearer the footpoints, with a wide spread across the loop ($\approx 50$ Mm).  \citet{to2024} suggested that these enhancements were due to reconnection downflows carrying low-FIP enhanced plasma onto the flare arcade, while evaporation brings chromospheric plasma upwards, leading to the observed distribution.    

Modeling work has recently begun to examine the role of evaporation on abundances \citep{reep2024}.  Depending on the assumed initial distribution of low-FIP material, strong evaporation can produce loops of photospheric composition with a narrow enhancement of low-FIP material near the apex \citep{benavitz2025}, or an enhancement that spreads across the whole loop \citep{reep2026}.  A major assumption of those works is that the low-FIP material was already enhanced in either the corona \citep{benavitz2025} or near the footpoints \citep{reep2026}.  Furthermore, with a strong electron beam, the evaporative flows act to strongly compress the low-FIP material, generally producing only a \textit{narrow} enhancement.  

In this work, we test the suggestion of \citet{to2024}: can directly injected material produce chemically enhanced post-flare arcades?  What are the resultant properties of the loops? We derive equations to inject material of arbitrary density, temperature, and abundance factor into a coronal loop.  We examine the effects of this direct injection both with and without simultaneous injection of a strong electron beam.  We synthesize both Ca/Ar and Fe/S ratios using full non-equilibrium ionization calculations.  We show that the observations in the gradual phase are consistent with direct injection of coronally-enhanced material, \textit{without} an electron beam or other strong chromospheric heating.

\section{Observations}
\label{sec:obs}

We first re-examine the distribution of the abundance factor in the 2017 September 10 flare, as measured by Hinode's Extreme ultraviolet Imaging Spectrometer (EIS; \citealt{culhane2007}).  The observations in \citet{to2024} focused on the gradual phase emission, showing measurements of the abundance factor at 18:39 (Figure 3 of that paper), more than 2 hours after the peak of the event.  However, it is interesting to consider whether and how this distribution evolves in time, so we briefly revisit here measurements in the other rasters at various times to see how the spatial distribution evolves in time.  

\begin{figure*}
    \centering
    \includegraphics[width=\linewidth]{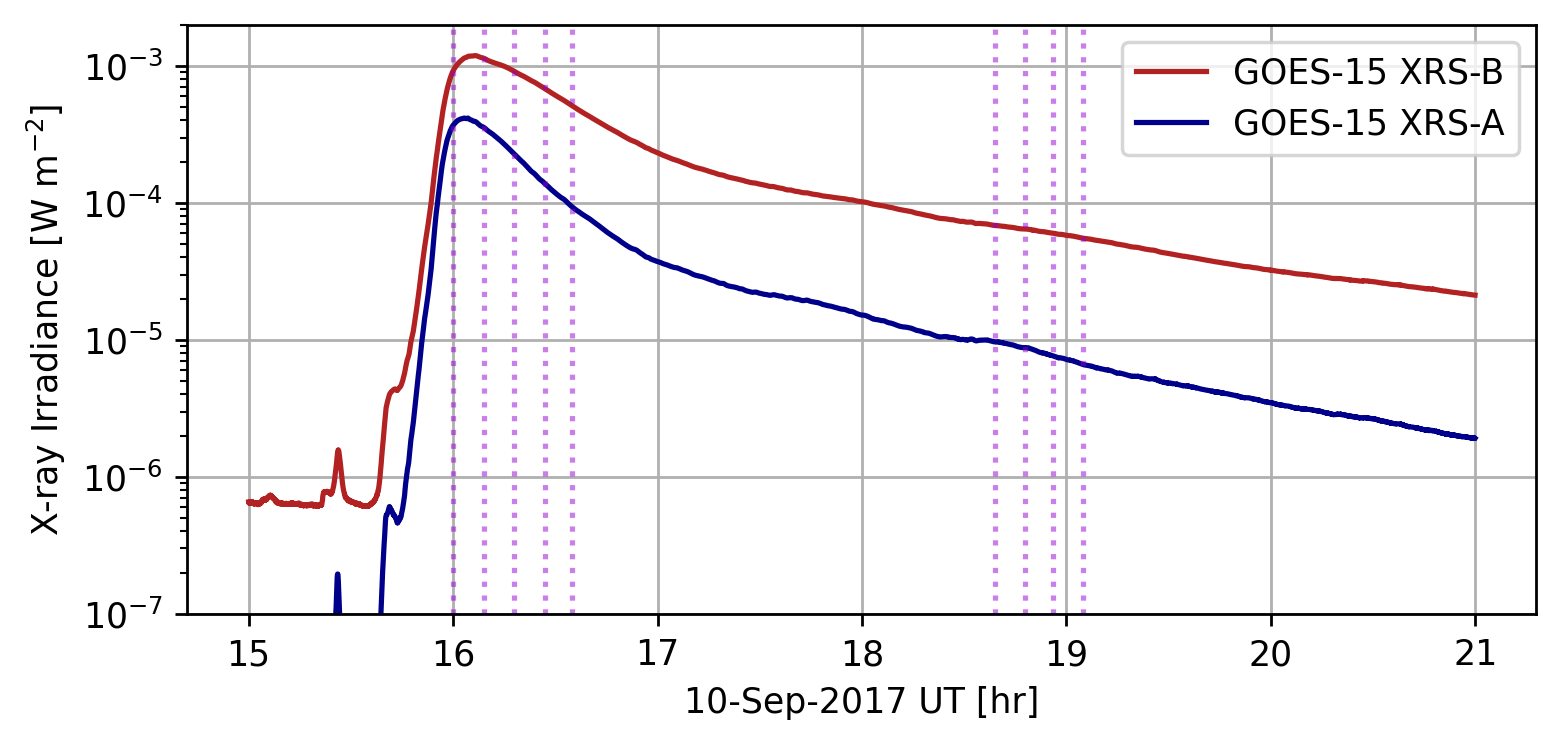}
    \includegraphics[width=\linewidth]{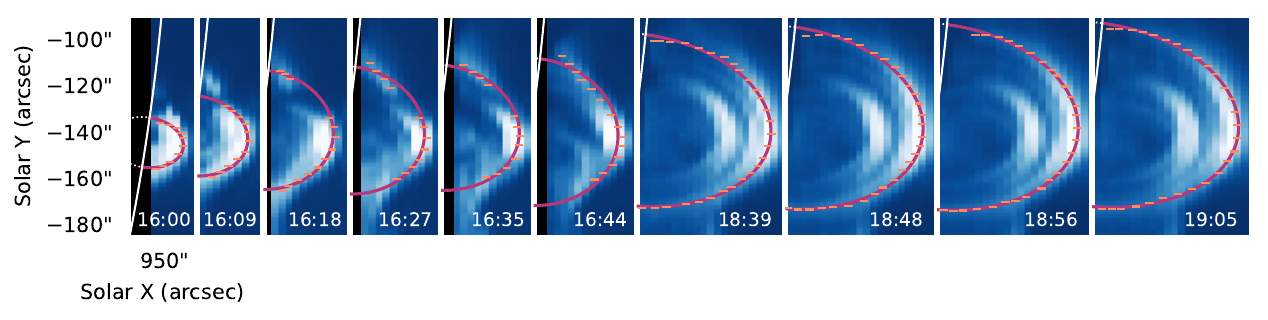}
    \caption{The GOES-15 XRS lightcurves and Hinode/EIS rasters for the beginning of the 10 September 2017 event.  The times of the 10 Hinode/EIS rasters are marked as dotted purple vertical lines.  The first raster occurred during the X-ray peak, the second just after the peak, while the others occurred throughout the gradual phase.  The scans of the EIS rasters are shown at bottom, with the red curve roughly tracing the the top of the arcade.  The orange ticks along the red curves mark the locations where we measure the abundance factor at each time.}
    \label{fig:obs_intensity}
\end{figure*}

The top panel of Figure \ref{fig:obs_intensity} shows the GOES-15 XRS lightcurves for the event, beginning around 15 UT.  The vertical dotted lines mark the times of the ten EIS rasters that scanned the arcade of the event, which occurred off the Western limb.  The first raster, at 16 UT, occurred during the X-ray peak, and the second just after the peak.  The remainder were taken at various times throughout the gradual phase.  The bottom panel shows that evolution of the flare arcade with time, as measured by EIS in \ion{Fe}{16} 262.984 \AA.  All 10 of the EIS rasters are shown sequentially, from left to right.  The orange ticks along the red curves indicate locations where we measure the abundance factor, corresponding to the tallest loop in the arcade at any given time.  As the flare evolves, the arcade grows in height and therefore the loops become significantly longer.  At most times, there are bright points seen near the apex of the arcade, sometimes referred to in the literature (\textit{e.g.} \citealt{doschek1996}) as bright ``knots'' of emission.  These bright points correspond to the locations of enhanced low-FIP plasma, as examined by \citet{to2024}.  

\begin{figure*}
    \centering
    \includegraphics[width=\linewidth]{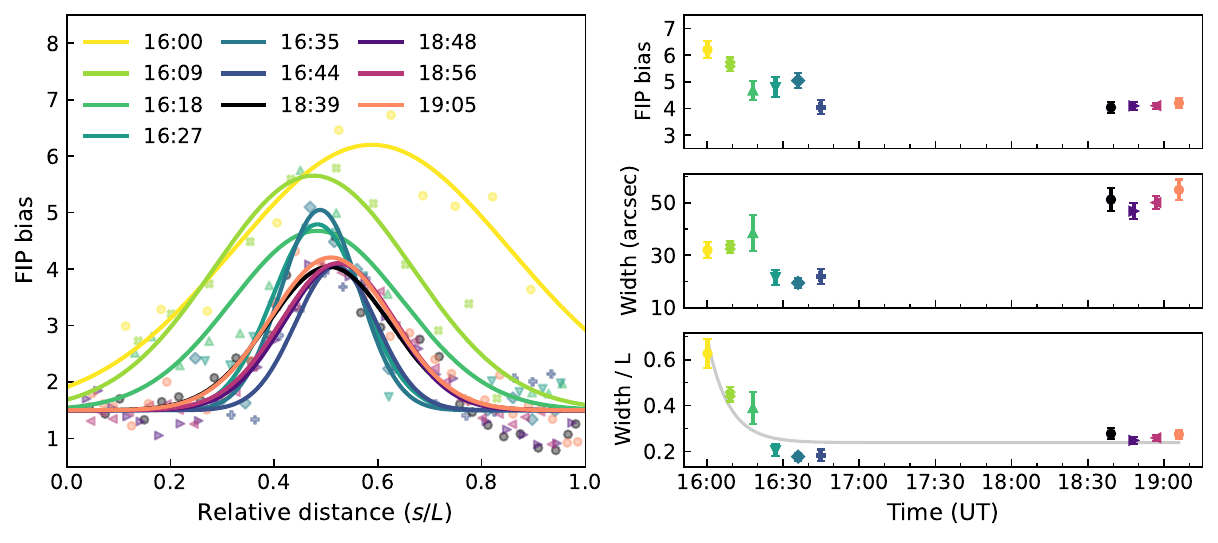}
    \caption{Measurements of the FIP bias (abundance factor) with the \ion{Ca}{14}/\ion{Ar}{14} line ratio near along the flare arcade shown in Figure \ref{fig:obs_intensity}.  The measured FIP biases are fitted with a Gaussian with background at 1.5.  The X-axis shows the relative distance, where the position is normalized by the fitted loop length at each time.  The plots at right, from top to bottom, show the fitted FIP bias amplitudes, FWHM of the fitted Gaussian, and that width divided by the total loop length. The grey line shows an exponential fit.
    }
    \label{fig:obs_abundance}
\end{figure*}

Using those Hinode/EIS rasters, we measure the relative FIP bias along the flare arcade with the \ion{Ca}{14} 193.874 \AA/\ion{Ar}{14} 194.396 \AA\ diagnostic. We follow the same method as in previous studies, using the EIS differential emission measure to minimize temperature and density effects \citep{brooks2010,hannah2012}. Figure \ref{fig:obs_abundance} shows these measurements along the arcade and in each EIS raster.  The plot at left shows the measurements of the FIP bias at each time along the red curves shown in Figure \ref{fig:obs_intensity}.  The distance is normalized to the length of the loop at each time.  The points mark the individual measurements, while the curves are Gaussian fits to the distribution, using a background level of 1.5.  The plots at right show how the distributions change in time: the amplitude of the FIP bias, the full-width-at-half-maximum (FWHM) of the Gaussian fit (in arcsec), and the width normalized to the loop length.  We see that the magnitude of the peak FIP bias decreases from a value of about 6 in the impulsive phase to around 3 in the later parts of the gradual phase.  The width of the region with enhanced FIP bias generally increases with time, initially around 30 arcsec (22 Mm), decreasing slightly to around 20 arcsec (14 Mm) during the early gradual phase, and then increasing to around 50 arcsec (36 Mm) at later times.  However, relative to the length of the loop, the portion of the enhanced region decreases approximately exponentially from around 60\% to 20\%, where it remains during the gradual phase.

From these observations, it is clear that the FIP bias evolves in both space and time.  The cause of these changes, however, is not readily apparent from the observations.  In previous modeling work, we examined the role of chromospheric evaporation.  If the corona is initially fully fractionated, the distribution becomes very sharply peaked near the apex of the loop, perhaps only a few percent of the loop length \citep{benavitz2025}.  If, on the other hand, the enhanced material is carried from the chromosphere, the distribution can be more widely distributed, but typically not sharply peaked (see the different cases in \citealt{reep2026}).  We examine a different possibility here: the direct injection of enhanced material from the reconnection outflow.   

\section{Equations}
\label{sec:eqns}

We model the direct injection of material from a reconnection site onto a coronal loop with the \texttt{HYDRAD} code \citep{bradshaw2003,bradshaw2013}.  The code solves the equations describing the conservation of mass, momentum, electron energy, and hydrogen energy.  Following a modification to the code presented in \citet{benavitz2025}, we also solve for the spatiotemporal variation of abundances of low-FIP elements, using an abundance factor $f$ relative to hydrogen to describe this variation.  We assume that all the low-FIP elements vary together for simplicity, and that their motion is dictated only by the bulk flow velocity $v$.  Although these are all trace elements, the optically thin radiative loss rate is calculated using the local abundances, so that this variation in $f$ feeds back into the electron energy equation.  As shown by \citet{benavitz2025}, this can directly cause the formation of coronal condensations, otherwise absent from impulsively-heated loop simulations \citep{reep2020}.  

We modify the continuity equations for both the bulk material and for low-FIP material by introducing a source term, a direct injection of material into the plasma.  We assume that the injection occurs over a Gaussian spatial distribution.  We abbreviate partial derivatives as $\frac{\partial}{\partial x} = \partial_{x}$.  The bulk continuity equation is then modified as
\begin{align}
    \partial_{t}(\rho) = -\frac{1}{A} \partial_{s}(\rho A v) + \frac{\rho_{in} w}{\sigma \sqrt{2\pi}} \exp{\Big(-\frac{(s-s_{0})^{2}}{2\sigma^{2}}\Big)}
\end{align}
\noindent where $\rho_{in}$ is the injected mass density, $w$ is the speed of the injection (transverse to the flux tube coordinate $s$), and $s_{0}$ is the position along the loop where we inject the mass over a Gaussian width $\sigma$.  

We allow the injected mass to have an arbitrary abundance factor $f_{in}$, and therefore also add a source term to our low-FIP continuity equation:
\begin{align}
    \partial_{t}(\rho f) = &-\frac{1}{A} \partial_{s}(\rho A v f) \nonumber \\ &+ \frac{\rho_{in} f_{in} w }{\sigma \sqrt{2\pi}} \exp{\Big(-\frac{(s-s_{0})^{2}}{2\sigma^{2}}\Big)}
\end{align}
\noindent As in \citet{benavitz2025}, we can recast this second equation simply in terms of the abundance factor by combining with the first equation to find:
\begin{align}
    \partial_{t} f = - v\ \partial_{s} f + \frac{(f_{in} - f)\rho_{in} w }{\rho\  \sigma \sqrt{2\pi}} \exp{\Big(-\frac{(s-s_{0})^{2}}{2\sigma^{2}}\Big)}
\end{align}
\noindent Note that if the injected material has the same abundance factor as the local plasma, \textit{i.e.} $f_{in} = f$, the injection does not affect the abundance factor, but does change the local mass density.

The momentum equation in \texttt{HYDRAD} is solved in a conservative form, and therefore does not require modification.  Specifically, since we assume the injection occurs at a speed $w$ transverse to the flux tube, it does not directly inject momentum parallel to the flux tube.  It does, however, affect the parallel flow speed, which can be seen simply.  If the source term in the bulk continuity equation is called $S_{in}$, then we can use the chain rule to show:
\begin{align}
    \partial_{t} (\rho v) + \frac{1}{A}\partial_{s}(A\rho v^{2}) = &\rho \Big(\partial_{t} v + v\ \partial_{s} v\Big ) \nonumber \\
    &+ v \Big(\partial_{t} \rho + \frac{1}{A} \partial_{s}(\rho A v) \Big) \nonumber \\ 
    = &\rho \Big(\partial_{t} v + v\ \partial_{s} v\Big ) + vS_{in}
\end{align}
\noindent which depends directly on the injected mass.  Effectively, adding mass slows down any existing flows.  In practice, the flows are close to zero during the time period over which we inject material.  

The energy equations require the addition of source terms.  That is, if we inject mass, we are also injecting internal energy into the plasma locally.  We assume that the injected plasma has a temperature $T_{in}$, and then the internal energy added to the electron and hydrogen energy equations is given by
\begin{align}
    \dot{\epsilon}_{e} &= \frac{3}{2} k_{B} T_{e,in} \Big(\frac{1.16}{1.32m_{H}}\Big) \frac{\rho_{in} w}{\sigma \sqrt{2\pi}}\exp{\Big(-\frac{(s-s_{0})^{2}}{2\sigma^{2}}\Big)}\\
    \dot{\epsilon}_{H} &= \frac{3}{2} k_{B} T_{H,in} \Big(\frac{1}{1.32m_{H}}\Big) \frac{\rho_{in} w}{\sigma \sqrt{2\pi}}\exp{\Big(-\frac{(s-s_{0})^{2}}{2\sigma^{2}}\Big)}
\end{align}
\noindent where the injected number densities are determined by the injected mass density.  We assume that the mass is dominated by hydrogen and helium, $\rho \approx \rho_{H} + \rho_{He} = \rho_{H} (1 + 4 A_{He}) \approx 1.32 \rho_{H}$, where $A_{He}$ is the abundance of helium.  We can therefore write $n_{H,in} = \frac{\rho_{in}}{1.32 m_{H}}$.  Since we inject at high temperatures ($T_{in} \gtrsim 10$ MK), we also assume that the electron number density is dominated by fully ionized hydrogen and helium, so $n_{e} \approx n_{H} + 2 n_{He} = n_{H} (1 + 2 A_{He}) \approx 1.16 n_{H}$.

\section{Simulations}
\label{sec:sims}

We focus on a qualitative comparison to the observational measurements of abundances in the 10 September 2017 X12 event.  In this flare, the current sheet was nearly isothermal at around 15 to 20 MK, with a density of around $n \approx 10^{10}$ cm$^{-3}$ \citep{warren2018}.  We assume that the injected temperatures are equilibrated: $T_{e,in} = T_{H,in}$.  Downflows onto the arcade occurred with flow speeds of a few hundred km s$^{-1}$, perhaps up to 900 km s$^{-1}$ \citep{yu2020}.  We also assume that the directly injected material has an abundance factor $f_{in} = 5$ to test the scenario presented in \citet{to2024}.  In all cases, we simulate loops of total length $2L = 70$ Mm, roughly corresponding to the ``upper loop'' examined by \citet{to2024}.  

We consider two cases:
\begin{enumerate}
    \item With photospheric abundances everywhere initially (\textit{i.e.} $f(s,t=0)=1)$, we inject material with coronal abundances near the apex.  
    \item With photospheric abundances everywhere initially, we inject material with coronal abundances near the apex while simultaneously initiating heating by a strong electron beam.  
\end{enumerate}

We examine these cases to understand if they can produce a similar distribution of abundance factors as reported by \citet{to2024}.  To do so, we show both the abundance factor $f$ of the simulation, as well as ``deriving'' an abundance factor with the same line ratios as in that paper (\ion{Ca}{14} 193.87 \AA\ to \ion{Ar}{14} 194.40 \AA\ and \ion{Fe}{16} 262.98 \AA\ to \ion{S}{13} 256.69 \AA).  We calculate the ionization fractions of these four elements with a full non-equilibrium ionization treatment \citep{bradshaw2004,bradshaw2013b}.

\subsection{Direct injection without direct chromospheric heating}
\label{subsec:direct}

\begin{figure*}
    \centering
    \includegraphics[width=\linewidth]{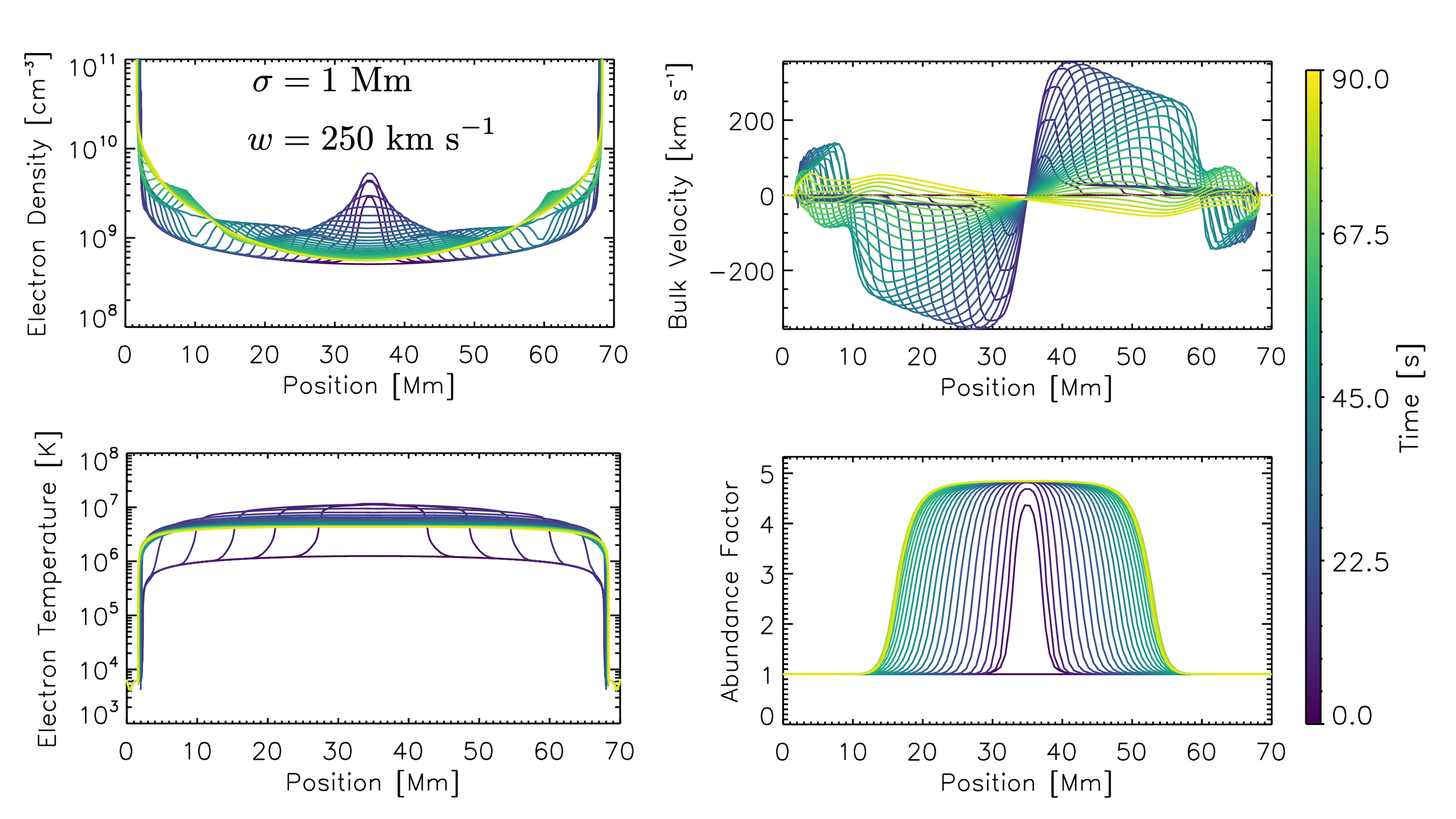}
    \caption{An example of direct injection at the apex onto a coronal loop, with $\sigma = 1$ Mm and $w = 250$ km s$^{-1}$.  From top left, the plots show the electron density, bulk velocity (flows to the right are positive), electron temperature, and abundance factor along the loop, at a 3 s cadence for the first 90 s after injection.  The initial injection causes the density, temperature, and abundance factor to rise sharply near the apex.  The increase in pressure causes flows to push material away from the apex, while conduction carries heat to the chromosphere that induces evaporation into the corona.  The abundance factor rises to $f=5$ near the apex, and spreads with the initial outward flows.  As the pressure at the apex decreases, the evaporative flows begin to dominate and the spread of $f$ ceases (around 70 s), beginning to compress back towards the apex. }
    \label{fig:uni_inj_s1v250}
\end{figure*}

We first show an example of a loop with uniform cross-sectional area in Figure \ref{fig:uni_inj_s1v250}.  We directly inject material for 10 s at the apex with $f_{in}=5$, at a density $n_H \approx 10^{10}$ cm$^{-3}$, with a transverse velocity $w = 250$ km s$^{-1}$, over a Gaussian width $\sigma = 1$ Mm.  The plots show the electron density, bulk velocity, electron temperature, and abundance factor along the loop (x-axis), as a function of time (line color, 3 s cadence, from 0 to 90 s).  

Initially, the injection of hot, dense material quickly causes the density to rise to $\approx 6 \times 10^{9}$ cm$^{-3}$, the temperature to $\approx 10$ MK, and the abundance factor to 5.  Because both the temperature and density rise, the gas pressure ($P \propto n T$) rises as well, creating a pressure gradient that drives flows and material away from the apex.  At the same time, thermal conduction begins to carry energy across the length of the corona, taking less than 20 s to reach and heat the chromosphere, inducing chromospheric evaporation.  Conduction smooths out the temperature gradient in the corona, and the flows moving away from the apex smooth the density gradient.  The pressure gradient at the apex therefore slowly dissipates until evaporation comes to dominate the flows, around 70 s into the simulation.  The abundance factor, initially increasing near the apex to around 5, is carried away from the apex with the flows, causing it to spread to a width of around 40 Mm.  As the flows reverse in direction, this begins to cause $f$ to compress back towards the apex.  

\begin{figure*}
    \centering
    \includegraphics[width=\linewidth]{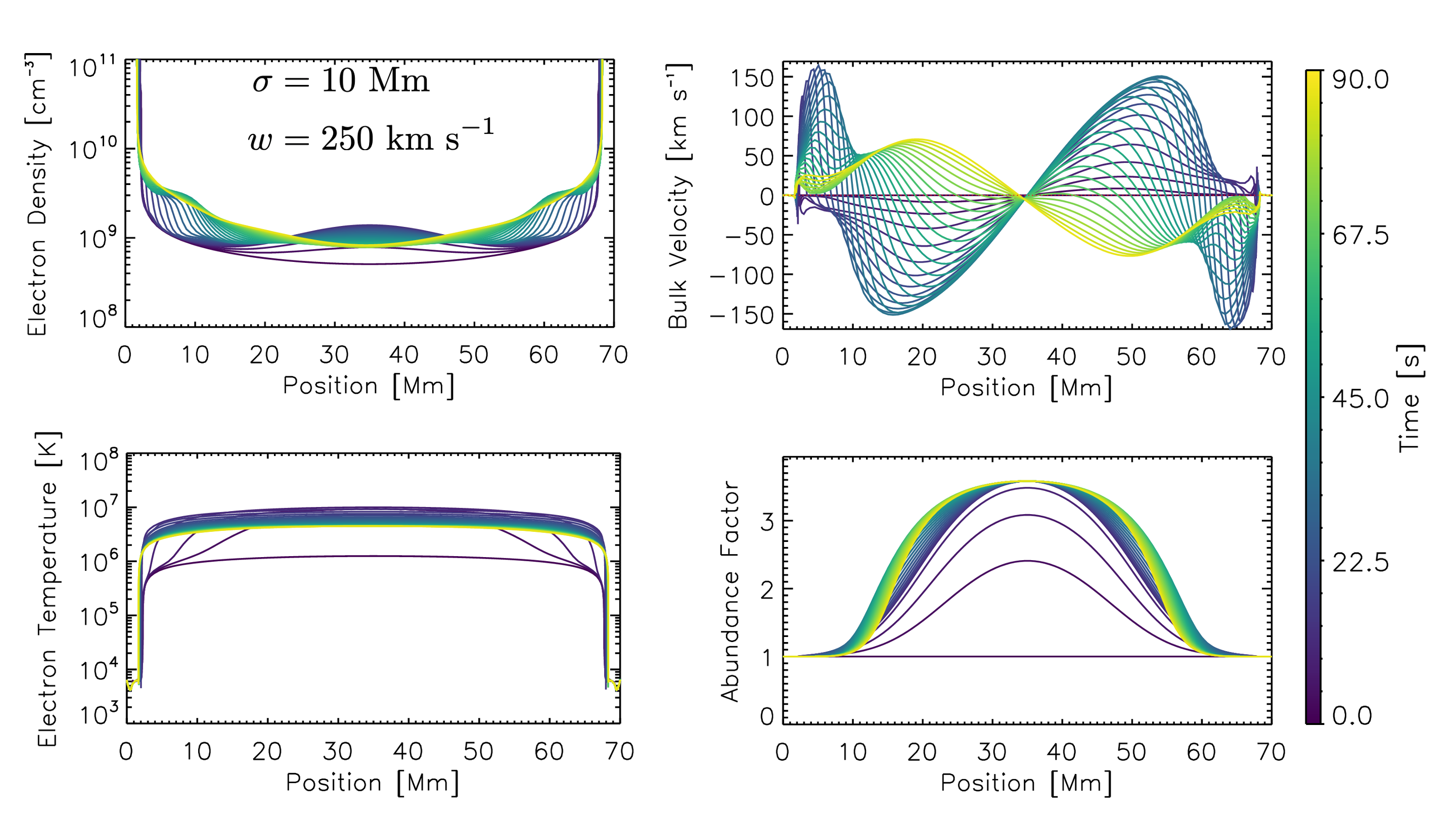}
    \caption{Similar to Figure \ref{fig:uni_inj_s1v250}, with $\sigma = 10$ Mm and $w = 250$ km s$^{-1}$.  The injection occurs over a much wider section of the loop, so that the initial rise in density and abundance factor are not as sharp.  The pressure increase is slightly less drastic, so the flow speeds are slower, but the mass is able to spread further from the injection site.  As before, the upflows from chromospheric evaporation eventually cause the material to stop spreading.}
    \label{fig:uni_inj_s10v250}
\end{figure*}

Contrast this with a case where the injection occurs over a wider section of the coronal loop, $\sigma = 10$ Mm, and otherwise equal parameters, shown in Figure \ref{fig:uni_inj_s10v250}.  In this case, the density and abundance factor rise near the apex, but are spread across a large section of the loop.  Because of this spread, the density only rises to a more modest $\approx 2 \times 10^{9}$ cm$^{-3}$.  The pressure increase near the apex drives a flow down the loop, but in this case, at slower speeds.  Simultaneously, the coronal temperature increases to almost 10 MK, and conduction begins to carry energy towards the chromosphere, which induces weak chromospheric evaporation.  As the pressure gradient near the apex of the loop reduces and the flows away from the apex weaken, evaporation begins to dominate, and the spread of the injected material ceases.  The abundance factor $f$ only increases to about 3.5 at the apex, but spreads across more than 50 Mm of the loop.  

\begin{figure*}
    \centering
    \includegraphics[width=\linewidth]{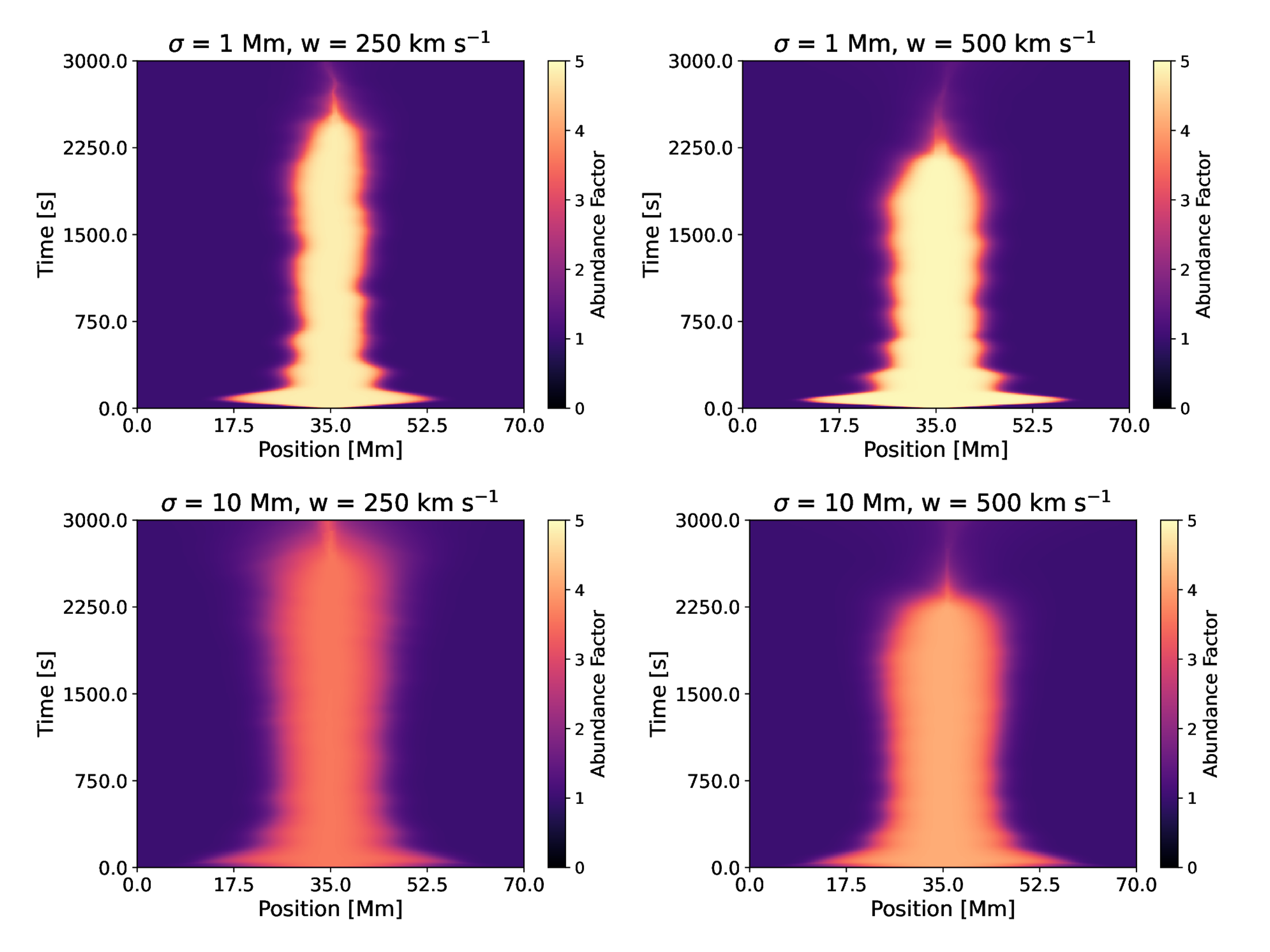}
    \caption{The long term behavior of the abundance factor $f$ in four loops of uniform area with directly injected mass, for different values of $\sigma$ and $w$.  A narrow injection results in a higher $f$ near the apex, that becomes relatively compressed.  A wider injection results in a lower $f$ that is relatively spread out.  A larger transverse speed $w$ increases the amount and rate at which mass and internal energy are injected.}
    \label{fig:uni_inj_3d}
\end{figure*}

We next examine how the abundance factor changes as the loop cools.  Figure \ref{fig:uni_inj_3d} shows the abundance factor for four cases, each permutation of $\sigma = [1, 10]$ Mm and $w = [250, 500]$ km s$^{-1}$.  The plots shows how $f$ evolves along the loop (x-axis) over time (y-axis).  The initial injection deposits mass and energy at the apex of the loop in all cases.  With a larger transverse velocity $w$, the injection rate is much higher, $\partial_t f \propto w$, so the cases with $w = 500$ km s$^{-1}$ rise quickly to $f=5$, while the others only increase to 3 -- 4.  Early on, the pressure at the apex rises, driving flows away from the apex that cause the low-FIP material to spread across the length of the corona.  As in the previous figures, when this pressure gradient dissipates, chromospheric evaporation begins to push the material back towards the apex.  This material remains centered around the apex until the loops collapse, forming a coronal condensation near the apex in all cases.  Critically, the width of the enhancement in $f$ depends on both the initial width of injection $\sigma$ and the transverse velocity $w$.  In all cases, at late times, \textit{e.g.} $ t = 1500$ s, the width of the low-FIP-enhanced region is $\approx$ 15 to 20 Mm, about 20\% to 30\% of the total loop length, broadly consistent with the gradual phase measurements in Figure \ref{fig:obs_abundance}.

\subsection{Direct injection along with direct chromospheric heating}
\label{subsec:directheating}

We next consider the case where we have both a direct injection of thermal plasma as well as an injected beam of non-thermal electrons.  We assume that the duration of injection is the same for both, 10 s.  The electron beam is assumed to carry an energy flux $F = 10^{11}$ erg s$^{-1}$ cm$^{-2}$, with a low energy cut-off $E_{c} = 20$ keV, spectral index $\delta = 5$, and zero pitch angle.  It is injected symmetrically to both legs of the loop.

\begin{figure*}
    \centering
    \includegraphics[width=\linewidth]{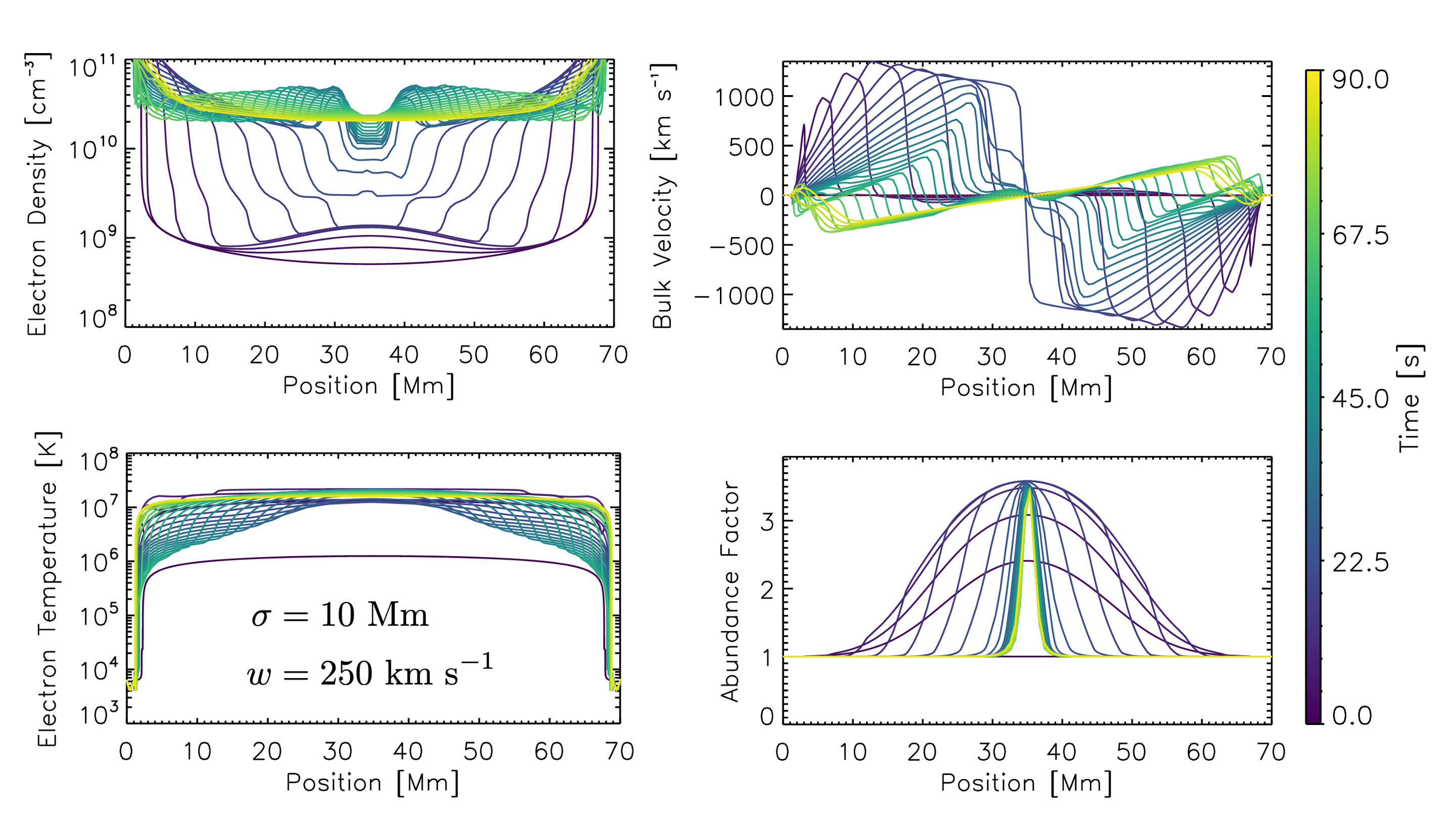}
    \caption{Similar to Figure \ref{fig:uni_inj_s10v250}, with the same parameters but the addition of a strong non-thermal electron beam.  The influence of the beam dominates the evolution of the loop.  The deposition of energy in the chromosphere drives a strong evaporative flow, exceeding 1,000 km s$^{-1}$ into the corona, that sharply increases the coronal density and pushes the low-FIP enhanced plasma to a very narrow region near the apex.}
    \label{fig:uni_bminj_s10v250}
\end{figure*}

We first examine a case, shown in Figure \ref{fig:uni_bminj_s10v250}, with identical injection parameters to the simulation shown in Figure \ref{fig:uni_inj_s10v250}, but the addition of the electron beam.  As before, the directly injected thermal plasma causes the temperature and density near the apex to rise modestly, and the abundance factor rises to around 3.5.  However, the beam deposits its energy via Coulomb collisions in the chromosphere \citep{emslie1978}, raising the pressure there and inducing explosive chromospheric evaporation.  In less than 10 seconds, the flows into the corona exceed 1,000 km s$^{-1}$, which quickly raises the coronal density to $\approx 3 \times 10^{10}$ cm$^{-3}$ and temperature to $\approx 20$ MK.  The (symmetric) flows push the enhanced low-FIP plasma towards the apex, where it becomes sharply peaked, similar to the beam-driven cases presented in \citet{benavitz2025}.  The width of this enhancement of the abundance factor is approximately 5 Mm, much narrower than measured at any time in Section \ref{sec:obs}.  

\begin{figure*}
    \centering
    \includegraphics[width=\linewidth]{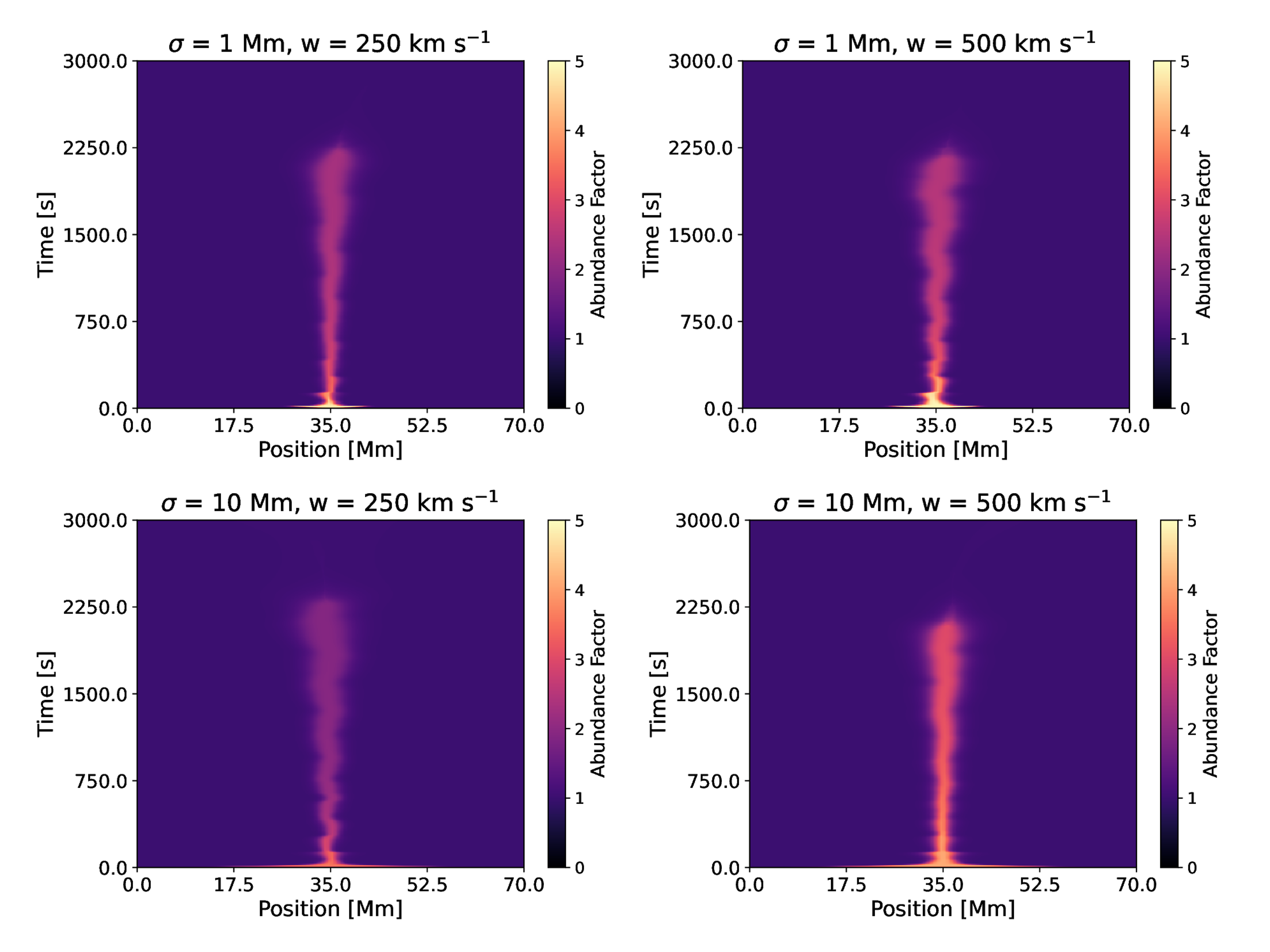}
    \caption{Similar to Figure \ref{fig:uni_inj_3d}, showing evolution of the abundance factor $f$ with the direct injection of material alongside a strong electron beam injection.  The electron beam induces strong evaporative flows that dominate the behavior, causing $f$ to be strongly constricted near the injection site, for all injection parameters and all times after the initial injection.  The magnitude of $f$, however, does depend on the injection parameters.}
    \label{fig:uni_bm_inj_3d}
\end{figure*}

We next consider different injection widths $\sigma$ and transverse velocities $w$.  In Figure \ref{fig:uni_bm_inj_3d}, we show the long-term behavior of the abundance factor in four simulations with both direct injection of material and an electron beam.  Initially, the directly injected mass is spread across a wide breadth, but the electron beam quickly induces chromospheric evaporation.  The flows from evaporation carry hot, dense material with $f=1$ into the corona from both legs of the loop.  These flows strongly constrict the location of enhanced low-FIP material, restricting it towards the apex of the loop in all cases.  After the evaporation ceases, only weak flows remain because the pressure gradients have dissipated, so the low-FIP enhanced plasma does not stray significantly from the apex.  As a consequence, at all times, the enhancement in $f$ is located in a narrow band of the corona, significantly more narrow than the cases without an electron beam.  The widths are $\approx$ 5 Mm, or around 7\% of the total loop length, narrower than any of the observed values in Figure \ref{fig:obs_abundance}.  When there is strong chromospheric heating (due to an electron beam or other mechanism), therefore, we might expect a relatively narrow enhancement in $f$.    

The observations presented in Section \ref{sec:obs} show relatively wide regions with enhanced $f$, particularly during the gradual phase.  This event was occulted, so hard X-ray observations of the footpoints are unavailable to determine the extent of electron beam heating.  Assuming the Neupert effect occurred in this event, however, the derivative of the soft X-ray emission can be used as a proxy for the hard X-rays, which suggests that electron acceleration ceased around UT16:10.  The width of the enhancement is seen to grow sharply between the EIS rasters at UT16:30 and UT19:00, perhaps suggesting that the strength of chromospheric heating reduced over time.  The observations of the late gradual phase are broadly consistent with the direct injection scenario and perhaps only weak chromospheric heating.  

\subsection{Forward Modeling of Line Ratios}
\label{subsec:forward}

One remaining question is whether the abundance factor $f(s,t)$ in the simulation corresponds to the FIP bias measurements presented in Section \ref{sec:obs} and \citet{to2024}.  In Hinode/EIS data, \ion{Ca}{14} 193.866 \AA\ to \ion{Ar}{14} 194.401 \AA\ and \ion{Fe}{16} 262.976 \AA\ to \ion{S}{13} 256.685 \AA\ were used in this event to estimate the relative abundances.  The \texttt{HYDRAD} simulations in this work solved for the non-equilibrium ionization states of all four of these elements, which we use to forward model the line emission to check the observational ratios.  We use the \texttt{fiasco} package \citep{barnes2026}, v0.8.1, with CHIANTI version 11.0.2 \citep{dufresne2024}, to calculate level populations and contribution functions necessary to estimate line intensities.  

\begin{figure*}
    \includegraphics[width=\linewidth]{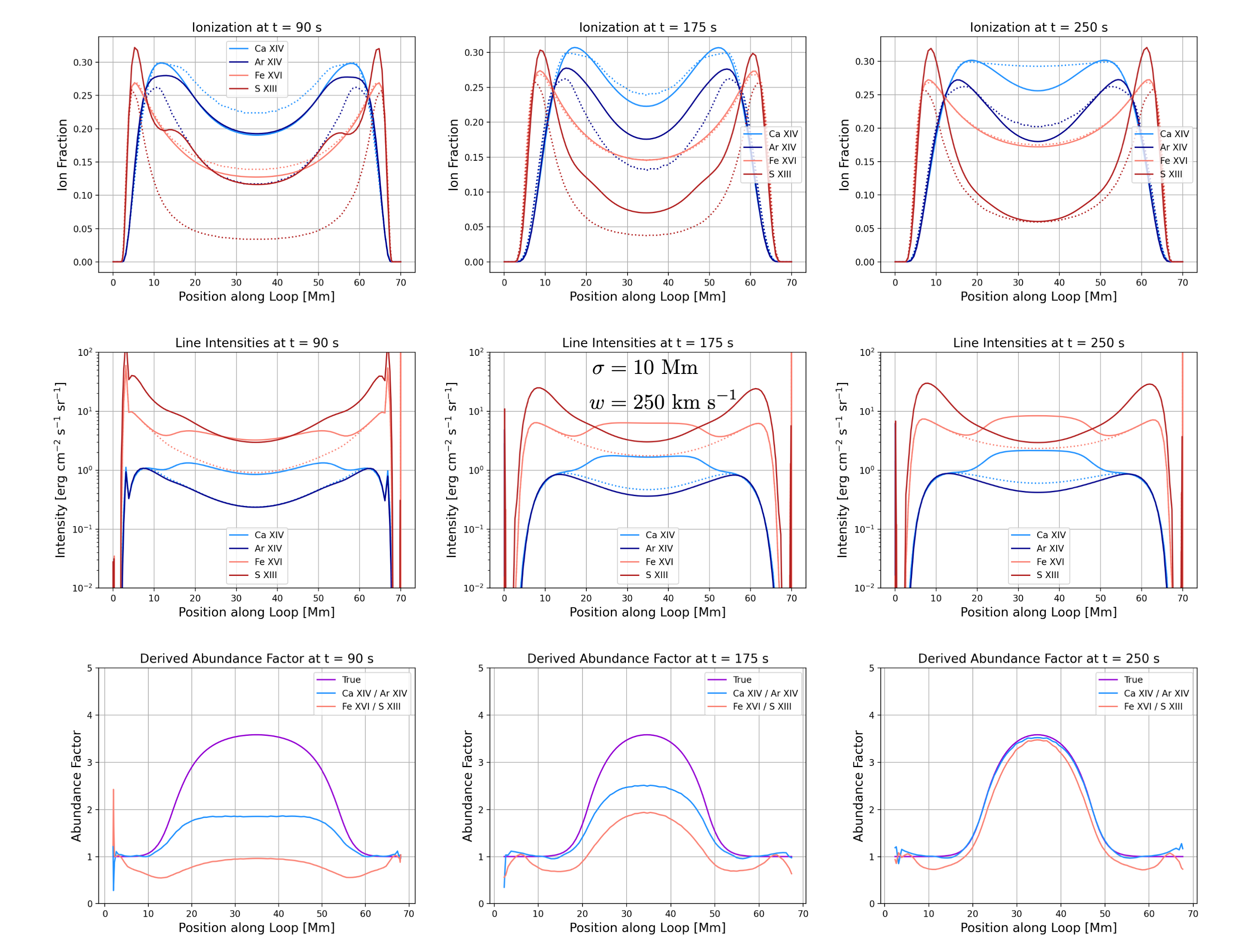}
    \caption{Synthetic abundance factors for the simulation presented in Figure \ref{fig:uni_inj_s10v250} with direct injection only, at select times (columns).  The top row shows the non-equilibrium (solid) and equilibrium (dotted) ionization fractions of the four ions along the loop, the middle row shows the line intensities, and the bottom row shows the derived abundance factors compared to the true value.  The dotted lines in the middle row show the low-FIP line intensities divided by the abundance factor.  The observational estimates of abundance factor $f$ generally reproduce the spatial distribution in the corona, but both ratios significantly underestimate the magnitude of $f$ when the ionization fraction departs from equilibrium.}
    \label{fig:forward}
\end{figure*}

Figure \ref{fig:forward} presents an estimate of observationally-derived abundance factors from these two line ratios.  We focus on the simulation presented in Figure \ref{fig:uni_inj_s10v250}, with $\sigma = 10$ Mm and $w = 250$ km s$^{-1}$, without an electron beam.  The columns each show one select time: 90 s, 175 s, and 250 s, respectively.  The top row shows the non-equilibrium ionization fractions (solid) of the four ions under consideration (for example, what fraction of Ca atoms are \ion{Ca}{14}), as compared to the equilibrium ionization fraction (dashed).  Early on, while the temperature is rapidly changing, these have departed from their equilibrium values, gradually tending towards equilibrium through collisions \citep{bradshaw2013b}.  The middle row shows the synthetic line intensities of the four spectral lines used in line ratios (solid lines).  For comparison, we also show the intensities of the \ion{Ca}{14} and \ion{Fe}{16} divided by the abundance factor $f(s)$ (dotted lines) to show how the varying abundance impacts the intensities.  Importantly, the intensities of the lines emitted by low-FIP ions are significantly brighter near the apex, and the localized brightenings persist as long as the abundance factor $f(s)$ is enhanced there.  This is consistent with the bright knots commonly seen near the looptops of flaring arcades, which are otherwise difficult to explain in hydrodynamic models because gradients in the temperature and density profiles become smoothed out \citep{doschek1996}.  In contrast, note that the dotted lines, $I(s) / f(s)$, show a relatively smooth variation across the corona, and are \textit{dimmest} at the apex.  

Finally, the bottom row shows the derived abundance factors from each line ratio compared to the true value in the simulation.  Early on, both ratios significantly underestimate the abundance factor because they depart from equilibrium ionization, but they gradually tend towards the true value.  On the other hand, the ratios generally reproduce the shape of the distribution at all times.  At late times, both ratios work well at measuring the abundance factor $f(s)$, suggesting that the observational measurements of the gradual phase were likely accurate.  However, how and when the ratios depart from the true abundance factor is an interesting question that merits further study, which will be the subject of forthcoming work.  Curiously, the distortion of the ratio due to non-equilibrium ionization can cause the apparent abundance factor to dip below 1 (in both ratios), appearing as though there is an inverse FIP effect, though the true $f$ never falls below 1.  


\section{Discussion}
\label{sec:disc}

\citet{doschek1996} raised an important question: ``What produces nonuniform emission in loops, such as the persistent bright knots at the tops of flare loops that are not predicted by numerical simulations?  Heat conduction and mass motions should smear out nonuniform emission within seconds.''  The intensity of spectral lines depends on the coronal density ($I \propto n_{i} n_{e}$), while the line width depends on the ion temperature ($\propto T_{i}$).  Pressure gradients induce flows that smooth out density gradients, while the diffusive process of thermal conduction ($F_{c} \propto \frac{\partial^{2} T}{\partial s^2}$) tends to smooth out gradients in temperature across the corona, particularly with the elevated temperatures reached in flares.  Previously, numerical simulations assumed elemental abundances fixed in both space and time, so that the predicted intensity would be nearly uniform across the corona, as noted by Doschek.  

In this work, we see how the ``bright knots'' might form: direct injection of low-FIP-enhanced plasma onto a flare arcade.  Many hydrodynamic models have relied solely on heating due to an electron beam, which drives chromospheric evaporation that heats and fills the loop(s).  For sufficiently strong beam heating rates, the temperatures and densities observed in a given flare can be reproduced, but gradients in these quantities become quickly smoothed, which would produce a (roughly) constant intensity across the corona.  A natural explanation of the bright knots is that there is a non-uniform distribution of elemental abundances.  For example, if the apex of the loop has a local enhancement in the density of Fe, the loop would appear locally brighter in spectral lines emitted by Fe.  Furthermore, gradients in the abundance of an element are not smoothed out in general (\textit{e.g.} Figure \ref{fig:uni_inj_3d}), so this local peak in brightness can persist for long periods of time, as observed.

\citet{to2024} additionally argued that the low-FIP-enhanced plasma is ``possibly confined by evaporation or ablation upflows from both footpoints.''  The modeling work here confirms this: chromospheric evaporation pushes inwards, confining the low-FIP-enhanced plasma to a relatively limited section of the loop.  The stronger the evaporation, the more limited in range the enhancement becomes (compare Figures \ref{fig:uni_inj_s10v250} and \ref{fig:uni_bminj_s10v250}).  Since we expect stronger evaporative flows during the impulsive phase of a flare, we therefore also expect the enhancement in abundance factor to be more narrowly peaked early in the flare.  While the region of low-FIP-enhanced plasma is indeed narrower early in the flare, it occupies a larger section of the loop, perhaps suggesting that either (1) evaporation is more limited than we expect, or (2) the evaporated plasma is \textit{also} low-FIP-enhanced.  Further observational work, particularly of the fractionation in the chromosphere, is necessary to disambiguate these scenarios.

We conclude from this work the following:
\begin{enumerate}
    \item The direct injection of material can produce a distribution of abundance factors $f(s)$ consistent with the observations of the 10 September 2017 flare, particularly during the gradual phase.  The increase in pressure near the injection site causes flows to develop, moving away from the apex, that gradually cause the width of the enhancement to grow, until chromospheric evaporation pushes the material back.
    \item Strong chromospheric evaporation, such as explosive evaporation induced by an electron beam, pushes material upwards into the corona.  If the evaporated material is photospheric in composition, this effectively limits the width of enhancements of $f(s)$ to a narrow region.  The impulsive phase observations of the 10 September 2017 event do indeed show a change in the width over the course of the flare, wider than the simulations with strong beam heating show.   
    \item Forward modeling of line ratios used to estimate $f(s)$ shows that they are roughly consistent with the true values, particularly at late times when the ionization fractions tend towards their equilibrium values, though non-equilibrium ionization can distort the observed amplitude of $f$.  The wide spatial distribution seen in the 2017 event is therefore likely an accurate measurement of the distribution of the abundance factor, though the magnitude of $f$ may be off due to non-equilibrium ionization effects.  This distribution is difficult to reconcile with strong evaporation, suggesting that the direct injection scenario is more plausible than strong chromospheric heating during the gradual phase.
    \item The bright knots that form near the top of solar flare arcades are consistent with localized enhancement of the abundance factor $f(s)$.  Spatial gradients $\frac{\partial f}{\partial s}$ are not smoothed by any diffusive process, unlike the temperature $T$ or density $n$.  As a result, intensity enhancements caused by peaks in $f$ can persist for a relatively long time.
    \item The spatiotemporal distribution of the abundance factor $f(s,t)$ is a strong and unique diagnostic of the energy release as well as mass transport.  It provides complementary diagnostics to the temperature, emission measure, density, or velocity, which are all regularly used to understand mass, momentum, and energy transport in solar flares and the quiescent corona.  High cadence observations with next-generation instruments like Solar-C EUVST \citep{shimizu2021} could use the abundance factor as a tracer of flows.  
\end{enumerate}

There remain many open questions.  One major question concerns the fractionation in the chromosphere: to what extent is the pre-flare atmosphere fractionated?  This drastically impacts the abundance distribution due to evaporation, but is currently poorly constained.  In some solar flares, there is even an inverse FIP effect, showing localized anomalous Ca/Ar line ratios \citep{doschek2015}, likely due to a depletion of Ca \citep{doschek2016}. It has been argued that evaporation of low-FIP-depleted plasma can explain these local variations \citep{baker2019}.  The inverse FIP effect is often seen in quiescent coronae of M-dwarfs \citep{laming2015}, and an inverse FIP effect can develop during flares on stars with a quiescent FIP bias \citep{nordon2008}.  The differences between these many cases warrants further examination.  

Two limitations of this work are (1) that we assume that low-FIP elements move with the bulk flows, and (2) that all low-FIP elements are treated the same here.  The ponderomotive force model often invoked to explain the FIP effect \citep{laming2015} requires that the low-FIP elements flow relative to the bulk flows.  To model this, either each species can be treated individually (\textit{e.g.} \citealt{martinez2023}) or a new momentum equation for low-FIP species must be introduced, which is the subject of forthcoming work.  Treating species individually has the advantage that the behavior should be more accurate, at the expense of significantly increased computational cost to solve more equations.

\acknowledgements{
The simulation data, including the code, configurations, and results, have been cached on Zenodo: https://doi.org/10.5281/zenodo.23487599
\\

A.S.H.T. acknowledges funding from the UPSaclaySTAR-Phy programme under the Marie Skłodowska-Curie Actions (MSCA COFUND), Grant Agreement No. 101216532.  Co-funded by the European Union. Views and opinions expressed are however those of the author(s) only and do not necessarily reflect those of the European Union or the REA. Neither the European Union nor the REA can be held responsible for them.
}

\nopagebreak

\leavevmode \newline

\clearpage
\onecolumngrid

\bibliography{apj}

@ARTICLE{baker2019,
       author = {{Baker}, Deborah and {van Driel-Gesztelyi}, Lidia and {Brooks}, David H. and {Valori}, Gherardo and {James}, Alexander W. and {Laming}, J. Martin and {Long}, David M. and {D{\'e}moulin}, Pascal and {Green}, Lucie M. and {Matthews}, Sarah A. and {Ol{\'a}h}, Katalin and {K{\H{o}}v{\'a}ri}, Zsolt},
        title = "{Transient Inverse-FIP Plasma Composition Evolution within a Solar Flare}",
      journal = {\apj},
         year = 2019,
        month = apr,
       volume = {875},
       number = {1},
          eid = {35},
        pages = {35},
          doi = {10.3847/1538-4357/ab07c1},
archivePrefix = {arXiv},
       eprint = {1902.06948},
 primaryClass = {astro-ph.SR},
       adsurl = {https://ui.adsabs.harvard.edu/abs/2019ApJ...875...35B}
}

@software{barnes2026,
   author       = {Will Barnes and
                   David Stansby and
                   Nick Murphy and
                   Jeffrey Reep and
                   Laura Hayes and
                   Stuart Mumford and
                   Nabil Freij},
   title        = {wtbarnes/fiasco},
   month        = mar,
   year         = 2026,
   publisher    = {Zenodo},
   version      = {0.8.1},
   doi          = {10.5281/zenodo.18942693},
   url          = {https://doi.org/10.5281/zenodo.18942693}
}

@ARTICLE{benavitz2025,
       author = {{Benavitz}, Luke Fushimi and {Reep}, Jeffrey W. and {Tarr}, Lucas A. and {To}, Andy S.~H.},
        title = "{Spatiotemporal Low First Ionization Potential Abundance: A Catalyst for Coronal Condensation}",
      journal = {\apj},
         year = 2025,
        month = oct,
       volume = {992},
       number = {1},
          eid = {4},
        pages = {4},
          doi = {10.3847/1538-4357/ae019d},
archivePrefix = {arXiv},
       eprint = {2504.05381},
 primaryClass = {astro-ph.SR},
       adsurl = {https://ui.adsabs.harvard.edu/abs/2025ApJ...992....4B}
}

@ARTICLE{bradshaw2003,
       author = {{Bradshaw}, S.~J. and {Mason}, H.~E.},
        title = "{A self-consistent treatment of radiation   in coronal loop modelling}",
      journal = {\aap},
         year = 2003,
        month = apr,
       volume = {401},
        pages = {699-709},
          doi = {10.1051/0004-6361:20030089},
       adsurl = {https://ui.adsabs.harvard.edu/abs/2003A&A...401..699B}
}

@ARTICLE{bradshaw2004,
       author = {{Bradshaw}, S.~J. and {Del Zanna}, G. and {Mason}, H.~E.},
        title = "{On the consequences of a non-equilibrium ionisation balance for compact flare emission and dynamics}",
      journal = {\aap},
         year = 2004,
        month = oct,
       volume = {425},
        pages = {287-299},
          doi = {10.1051/0004-6361:20040521},
       adsurl = {https://ui.adsabs.harvard.edu/abs/2004A&A...425..287B}
}

@ARTICLE{bradshaw2013,
       author = {{Bradshaw}, S.~J. and {Cargill}, P.~J.},
        title = "{The Influence of Numerical Resolution on Coronal Density in Hydrodynamic Models of Impulsive Heating}",
      journal = {\apj},
         year = 2013,
        month = jun,
       volume = {770},
       number = {1},
          eid = {12},
        pages = {12},
          doi = {10.1088/0004-637X/770/1/12},
archivePrefix = {arXiv},
       eprint = {1305.1902},
 primaryClass = {astro-ph.SR},
       adsurl = {https://ui.adsabs.harvard.edu/abs/2013ApJ...770...12B}
}

@ARTICLE{bradshaw2013b,
       author = {{Bradshaw}, Stephen J. and {Raymond}, John},
        title = "{Collisional and Radiative Processes in Optically Thin Plasmas}",
      journal = {\ssr},
         year = 2013,
        month = oct,
       volume = {178},
       number = {2-4},
        pages = {271-306},
          doi = {10.1007/s11214-013-9970-0},
archivePrefix = {arXiv},
       eprint = {1303.4468},
 primaryClass = {astro-ph.SR},
       adsurl = {https://ui.adsabs.harvard.edu/abs/2013SSRv..178..271B}
}

@article{brooks2010,
	author = {Brooks, David H. and Warren, Harry P.},
	title = {{ESTABLISHING A CONNECTION BETWEEN ACTIVE REGION OUTFLOWS AND THE SOLAR WIND: ABUNDANCE MEASUREMENTS WITH EIS/HINODE}},
	journal = {\apjl},
	volume = {727},
	number = {1},
	pages = {L13},
	year = {2011},
	month = {Jan},
	issn = {2041-8205},
	publisher = {IOP Publishing},
	doi = {10.1088/2041-8205/727/1/l13}
}

@article{brooks2026,
    author = {Brooks, David H. and Baker, Deborah and Long, David M. and Testa, Paola and Warren, Harry P.},
    title = {Evolution of solar and stellar coronal abundances owing to magnetic activity},
    journal = {Philosophical Transactions of the Royal Society A: Mathematical, Physical and Engineering Sciences},
    volume = {384},
    number = {2323},
    pages = {20250188},
    year = {2026},
    month = {07},
    issn = {1364-503X},
    doi = {10.1098/rsta.2025.0188},
    url = {https://doi.org/10.1098/rsta.2025.0188},
    eprint = {https://royalsocietypublishing.org/rsta/article-pdf/doi/10.1098/rsta.2025.0188/6149771/rsta20250188.pdf},
}

@INCOLLECTION{carmichael1964,
       author = {{Carmichael}, H.},
        title = "{A Process for Flares}",
    booktitle = {NASA Special Publication},
         year = 1964,
       editor = {{Hess}, Wilmot N.},
       volume = {50},
        pages = {451},
       adsurl = {https://ui.adsabs.harvard.edu/abs/1964NASSP..50..451C}
}

@ARTICLE{cheng2018,
       author = {{Cheng}, X. and {Li}, Y. and {Wan}, L.~F. and {Ding}, M.~D. and {Chen}, P.~F. and {Zhang}, J. and {Liu}, J.~J.},
        title = "{Observations of Turbulent Magnetic Reconnection within a Solar Current Sheet}",
      journal = {\apj},
         year = 2018,
        month = oct,
       volume = {866},
       number = {1},
          eid = {64},
        pages = {64},
          doi = {10.3847/1538-4357/aadd16},
archivePrefix = {arXiv},
       eprint = {1808.06071},
 primaryClass = {astro-ph.SR},
       adsurl = {https://ui.adsabs.harvard.edu/abs/2018ApJ...866...64C}
}

@ARTICLE{culhane2007,
       author = {{Culhane}, J.~L. and {Harra}, L.~K. and {James}, A.~M. and {Al-Janabi}, K. and {Bradley}, L.~J. and {Chaudry}, R.~A. and {Rees}, K. and {Tandy}, J.~A. and {Thomas}, P. and {Whillock}, M.~C.~R. and {Winter}, B. and {Doschek}, G.~A. and {Korendyke}, C.~M. and {Brown}, C.~M. and {Myers}, S. and {Mariska}, J. and {Seely}, J. and {Lang}, J. and {Kent}, B.~J. and {Shaughnessy}, B.~M. and {Young}, P.~R. and {Simnett}, G.~M. and {Castelli}, C.~M. and {Mahmoud}, S. and {Mapson-Menard}, H. and {Probyn}, B.~J. and {Thomas}, R.~J. and {Davila}, J. and {Dere}, K. and {Windt}, D. and {Shea}, J. and {Hagood}, R. and {Moye}, R. and {Hara}, H. and {Watanabe}, T. and {Matsuzaki}, K. and {Kosugi}, T. and {Hansteen}, V. and {Wikstol}, {\O}.},
        title = "{The EUV Imaging Spectrometer for Hinode}",
      journal = {\solphys},
         year = 2007,
        month = jun,
       volume = {243},
       number = {1},
        pages = {19-61},
          doi = {10.1007/s01007-007-0293-1},
       adsurl = {https://ui.adsabs.harvard.edu/abs/2007SoPh..243...19C}
}

@ARTICLE{dennis2015,
       author = {{Dennis}, Brian R. and {Phillips}, Kenneth J.~H. and {Schwartz}, Richard A. and {Tolbert}, Anne K. and {Starr}, Richard D. and {Nittler}, Larry R.},
        title = "{Solar Flare Element Abundances from the Solar Assembly for X-Rays (SAX) on MESSENGER}",
      journal = {\apj},
         year = 2015,
        month = apr,
       volume = {803},
       number = {2},
          eid = {67},
        pages = {67},
          doi = {10.1088/0004-637X/803/2/67},
archivePrefix = {arXiv},
       eprint = {1503.01602},
 primaryClass = {astro-ph.SR},
       adsurl = {https://ui.adsabs.harvard.edu/abs/2015ApJ...803...67D}
}

@ARTICLE{doschek1996,
       author = {{Doschek}, G.~A. and {Mariska}, John T. and {Sakao}, T.},
        title = "{Soft X-Ray Flare Dynamics}",
      journal = {\apj},
         year = 1996,
        month = mar,
       volume = {459},
        pages = {823},
          doi = {10.1086/176946},
       adsurl = {https://ui.adsabs.harvard.edu/abs/1996ApJ...459..823D}
}

@ARTICLE{doschek2014,
       author = {{Doschek}, G.~A. and {McKenzie}, D.~E. and {Warren}, H.~P.},
        title = "{Plasma Dynamics Above Solar Flare Soft X-Ray Loop Tops}",
      journal = {\apj},
         year = 2014,
        month = jun,
       volume = {788},
       number = {1},
          eid = {26},
        pages = {26},
          doi = {10.1088/0004-637X/788/1/26},
       adsurl = {https://ui.adsabs.harvard.edu/abs/2014ApJ...788...26D}
}

@ARTICLE{doschek2015,
       author = {{Doschek}, G.~A. and {Warren}, H.~P. and {Feldman}, U.},
        title = "{Anomalous Relative Ar/Ca Coronal Abundances Observed by the Hinode/EUV Imaging Spectrometer Near Sunspots}",
      journal = {\apjl},
         year = 2015,
        month = jul,
       volume = {808},
       number = {1},
          eid = {L7},
        pages = {L7},
          doi = {10.1088/2041-8205/808/1/L7},
       adsurl = {https://ui.adsabs.harvard.edu/abs/2015ApJ...808L...7D}
}

@ARTICLE{doschek2016,
       author = {{Doschek}, G.~A. and {Warren}, H.~P.},
        title = "{The Mysterious Case of the Solar Argon Abundance near Sunspots in Flares}",
      journal = {\apj},
         year = 2016,
        month = jul,
       volume = {825},
       number = {1},
          eid = {36},
        pages = {36},
          doi = {10.3847/0004-637X/825/1/36},
       adsurl = {https://ui.adsabs.harvard.edu/abs/2016ApJ...825...36D}
}

@ARTICLE{doschek2018,
       author = {{Doschek}, G.~A. and {Warren}, H.~P. and {Harra}, L.~K. and {Culhane}, J.~L. and {Watanabe}, T. and {Hara}, H.},
        title = "{Photospheric and Coronal Abundances in an X8.3 Class Limb Flare}",
      journal = {\apj},
         year = 2018,
        month = feb,
       volume = {853},
       number = {2},
          eid = {178},
        pages = {178},
          doi = {10.3847/1538-4357/aaa4f5},
       adsurl = {https://ui.adsabs.harvard.edu/abs/2018ApJ...853..178D}
}

@ARTICLE{dufresne2024,
       author = {{Dufresne}, R.~P. and {Del Zanna}, G. and {Young}, P.~R. and {Dere}, K.~P. and {Deliporanidou}, E. and {Barnes}, W.~T. and {Landi}, E.},
        title = "{CHIANTI{\textemdash}An Atomic Database for Emission Lines{\textemdash}Paper. XVIII. Version 11, Advanced Ionization Equilibrium Models: Density and Charge Transfer Effects}",
      journal = {\apj},
         year = 2024,
        month = oct,
       volume = {974},
       number = {1},
          eid = {71},
        pages = {71},
          doi = {10.3847/1538-4357/ad6765},
archivePrefix = {arXiv},
       eprint = {2403.16922},
 primaryClass = {astro-ph.SR},
       adsurl = {https://ui.adsabs.harvard.edu/abs/2024ApJ...974...71D}
}

@ARTICLE{emslie1978,
       author = {{Emslie}, A.~G.},
        title = "{The collisional interaction of a beam of charged particles with a hydrogen target of arbitrary ionization level.}",
      journal = {\apj},
         year = 1978,
        month = aug,
       volume = {224},
        pages = {241-246},
          doi = {10.1086/156371},
       adsurl = {https://ui.adsabs.harvard.edu/abs/1978ApJ...224..241E}
}

@ARTICLE{feldman1998,
       author = {{Feldman}, U.},
        title = "{FIP Effect in the Solar Upper Atmosphere: Spectroscopic Results}",
      journal = {\ssr},
         year = 1998,
        month = may,
       volume = {85},
        pages = {227-240},
          doi = {10.1023/A:1005146332450},
       adsurl = {https://ui.adsabs.harvard.edu/abs/1998SSRv...85..227F}
}

@ARTICLE{feldman2000,
       author = {{Feldman}, U. and {Laming}, J.~M.},
        title = "{Element Abundances in the Upper Atmospheres of the Sun and Stars: Update of Observational Results}",
      journal = {\physscr},
         year = 2000,
        month = feb,
       volume = {61},
        pages = {222},
          doi = {10.1238/Physica.Regular.061a00222},
       adsurl = {https://ui.adsabs.harvard.edu/abs/2000PhyS...61..222F}
}

@ARTICLE{fludra1995,
       author = {{Fludra}, A. and {Schmelz}, J.~T.},
        title = "{Absolute Abundances of Flaring Coronal Plasma Derived from SMM Spectral Observations}",
      journal = {\apj},
         year = 1995,
        month = jul,
       volume = {447},
        pages = {936},
          doi = {10.1086/175931},
       adsurl = {https://ui.adsabs.harvard.edu/abs/1995ApJ...447..936F}
}

@ARTICLE{french2024,
       author = {{French}, Ryan J. and {Yu}, Sijie and {Chen}, Bin and {Shen}, Chengcai and {Matthews}, Sarah A.},
        title = "{Doppler signature of a possible termination shock in an off-limb solar flare}",
      journal = {\mnras},
         year = 2024,
        month = mar,
       volume = {528},
       number = {4},
        pages = {6836-6844},
          doi = {10.1093/mnras/stae430},
archivePrefix = {arXiv},
       eprint = {2402.04445},
 primaryClass = {astro-ph.SR},
       adsurl = {https://ui.adsabs.harvard.edu/abs/2024MNRAS.528.6836F}
}

@ARTICLE{french2025,
       author = {{French}, Ryan J. and {Kazachenko}, Maria D. and {Mihailescu}, Teodora and {Reeves}, Katharine K.},
        title = "{Spectroscopic Observations of Supra-arcade Downflows}",
      journal = {\apjl},
         year = 2025,
        month = jun,
       volume = {986},
       number = {1},
          eid = {L16},
        pages = {L16},
          doi = {10.3847/2041-8213/adde59},
archivePrefix = {arXiv},
       eprint = {2505.22624},
 primaryClass = {astro-ph.SR},
       adsurl = {https://ui.adsabs.harvard.edu/abs/2025ApJ...986L..16F}
}

@article{hannah2012,
	author = {Hannah, I. G. and Kontar, E. P.},
	title = {{Differential emission measures from the regularized inversion of Hinode and SDO data}},
	journal = {\aap},
	volume = {539},
	pages = {A146},
	year = {2012},
	month = mar,
	issn = {0004-6361},
	publisher = {EDP Sciences},
	doi = {10.1051/0004-6361/201117576}
}

@ARTICLE{hayes2019,
       author = {{Hayes}, Laura A. and {Gallagher}, Peter T. and {Dennis}, Brian R. and {Ireland}, Jack and {Inglis}, Andrew and {Morosan}, Diana E.},
        title = "{Persistent Quasi-periodic Pulsations during a Large X-class Solar Flare}",
      journal = {\apj},
         year = 2019,
        month = apr,
       volume = {875},
       number = {1},
          eid = {33},
        pages = {33},
          doi = {10.3847/1538-4357/ab0ca3},
archivePrefix = {arXiv},
       eprint = {1903.01328},
 primaryClass = {astro-ph.SR},
       adsurl = {https://ui.adsabs.harvard.edu/abs/2019ApJ...875...33H}
}

@ARTICLE{katsuda2020,
       author = {{Katsuda}, Satoru and {Ohno}, Masanori and {Mori}, Koji and {Beppu}, Tatsuhiko and {Kanemaru}, Yoshiaki and {Tashiro}, Makoto S. and {Terada}, Yukikatsu and {Sato}, Kosuke and {Morita}, Kae and {Sagara}, Hikari and {Ogawa}, Futa and {Takahashi}, Haruya and {Murakami}, Hiroshi and {Nobukawa}, Masayoshi and {Tsunemi}, Hiroshi and {Hayashida}, Kiyoshi and {Matsumoto}, Hironori and {Noda}, Hirofumi and {Nakajima}, Hiroshi and {Ezoe}, Yuichiro and {Tsuboi}, Yohko and {Maeda}, Yoshitomo and {Yokoyama}, Takaaki and {Narukage}, Noriyuki},
        title = "{Inverse First Ionization Potential Effects in Giant Solar Flares Found from Earth X-Ray Albedo with Suzaku/XIS}",
      journal = {\apj},
         year = 2020,
        month = mar,
       volume = {891},
       number = {2},
          eid = {126},
        pages = {126},
          doi = {10.3847/1538-4357/ab7207},
archivePrefix = {arXiv},
       eprint = {2001.10643},
 primaryClass = {astro-ph.SR},
       adsurl = {https://ui.adsabs.harvard.edu/abs/2020ApJ...891..126K}
}

@ARTICLE{kopp1976,
       author = {{Kopp}, R.~A. and {Pneuman}, G.~W.},
        title = "{Magnetic reconnection in the corona and the loop prominence phenomenon.}",
      journal = {\solphys},
         year = 1976,
        month = sep,
       volume = {50},
       number = {1},
        pages = {85-98},
          doi = {10.1007/BF00206193},
       adsurl = {https://ui.adsabs.harvard.edu/abs/1976SoPh...50...85K}
}

@ARTICLE{laming2004,
       author = {{Laming}, J. Martin},
        title = "{A Unified Picture of the First Ionization Potential and Inverse First Ionization Potential Effects}",
      journal = {\apj},
         year = 2004,
        month = oct,
       volume = {614},
       number = {2},
        pages = {1063-1072},
          doi = {10.1086/423780},
archivePrefix = {arXiv},
       eprint = {astro-ph/0405230},
 primaryClass = {astro-ph},
       adsurl = {https://ui.adsabs.harvard.edu/abs/2004ApJ...614.1063L}
}

@ARTICLE{laming2015,
       author = {{Laming}, J. Martin},
        title = "{The FIP and Inverse FIP Effects in Solar and Stellar Coronae}",
      journal = {Living Reviews in Solar Physics},
         year = 2015,
        month = dec,
       volume = {12},
       number = {1},
          eid = {2},
        pages = {2},
          doi = {10.1007/lrsp-2015-2},
archivePrefix = {arXiv},
       eprint = {1504.08325},
 primaryClass = {astro-ph.SR},
       adsurl = {https://ui.adsabs.harvard.edu/abs/2015LRSP...12....2L}
}

@ARTICLE{laming2021,
       author = {{Laming}, J. Martin},
        title = "{The FIP and Inverse-FIP Effects in Solar Flares}",
      journal = {\apj},
         year = 2021,
        month = mar,
       volume = {909},
       number = {1},
          eid = {17},
        pages = {17},
          doi = {10.3847/1538-4357/abd9c3},
archivePrefix = {arXiv},
       eprint = {2101.03038},
 primaryClass = {astro-ph.SR},
       adsurl = {https://ui.adsabs.harvard.edu/abs/2021ApJ...909...17L}
}

@ARTICLE{li2018,
       author = {{Li}, Y. and {Xue}, J.~C. and {Ding}, M.~D. and {Cheng}, X. and {Su}, Y. and {Feng}, L. and {Hong}, J. and {Li}, H. and {Gan}, W.~Q.},
        title = "{Spectroscopic Observations of a Current Sheet in a Solar Flare}",
      journal = {\apjl},
         year = 2018,
        month = jan,
       volume = {853},
       number = {1},
          eid = {L15},
        pages = {L15},
          doi = {10.3847/2041-8213/aaa6c0},
archivePrefix = {arXiv},
       eprint = {1801.03631},
 primaryClass = {astro-ph.SR},
       adsurl = {https://ui.adsabs.harvard.edu/abs/2018ApJ...853L..15L}
}

@ARTICLE{liu2013,
       author = {{Liu}, Wei and {Chen}, Qingrong and {Petrosian}, Vah{\'e}},
        title = "{Plasmoid Ejections and Loop Contractions in an Eruptive M7.7 Solar Flare: Evidence of Particle Acceleration and Heating in Magnetic Reconnection Outflows}",
      journal = {\apj},
         year = 2013,
        month = apr,
       volume = {767},
       number = {2},
          eid = {168},
        pages = {168},
          doi = {10.1088/0004-637X/767/2/168},
archivePrefix = {arXiv},
       eprint = {1303.3321},
 primaryClass = {astro-ph.SR},
       adsurl = {https://ui.adsabs.harvard.edu/abs/2013ApJ...767..168L}
}

@ARTICLE{longcope2018,
       author = {{Longcope}, Dana and {Unverferth}, John and {Klein}, Courtney and {McCarthy}, Marika and {Priest}, Eric},
        title = "{Evidence for Downflows in the Narrow Plasma Sheet of 2017 September 10 and Their Significance for Flare Reconnection}",
      journal = {\apj},
         year = 2018,
        month = dec,
       volume = {868},
       number = {2},
          eid = {148},
        pages = {148},
          doi = {10.3847/1538-4357/aaeac4},
       adsurl = {https://ui.adsabs.harvard.edu/abs/2018ApJ...868..148L}
}

@ARTICLE{martinez2023,
       author = {{Mart{\'\i}nez-Sykora}, Juan and {De Pontieu}, Bart and {Hansteen}, Viggo H. and {Testa}, Paola and {Wargnier}, Q.~M. and {Szydlarski}, Mikolaj},
        title = "{The Impact of Multifluid Effects in the Solar Chromosphere on the Ponderomotive Force under SE and NEQ Ionization Conditions}",
      journal = {\apj},
         year = 2023,
        month = jun,
       volume = {949},
       number = {2},
          eid = {112},
        pages = {112},
          doi = {10.3847/1538-4357/acc465},
archivePrefix = {arXiv},
       eprint = {2211.09361},
 primaryClass = {astro-ph.SR},
       adsurl = {https://ui.adsabs.harvard.edu/abs/2023ApJ...949..112M}
}

@ARTICLE{mckenzie1999,
       author = {{McKenzie}, D.~E. and {Hudson}, H.~S.},
        title = "{X-Ray Observations of Motions and Structure above a Solar Flare Arcade}",
      journal = {\apjl},
         year = 1999,
        month = jul,
       volume = {519},
       number = {1},
        pages = {L93-L96},
          doi = {10.1086/312110},
       adsurl = {https://ui.adsabs.harvard.edu/abs/1999ApJ...519L..93M}
}

@ARTICLE{mondal2021,
       author = {{Mondal}, Biswajit and {Sarkar}, Aveek and {Vadawale}, Santosh V. and {Mithun}, N.~P.~S. and {Janardhan}, P. and {Del Zanna}, Giulio and {Mason}, Helen E. and {Mitra-Kraev}, Urmila and {Narendranath}, S.},
        title = "{Evolution of Elemental Abundances during B-Class Solar Flares: Soft X-Ray Spectral Measurements with Chandrayaan-2 XSM}",
      journal = {\apj},
         year = 2021,
        month = oct,
       volume = {920},
       number = {1},
          eid = {4},
        pages = {4},
          doi = {10.3847/1538-4357/ac14c1},
archivePrefix = {arXiv},
       eprint = {2107.07825},
 primaryClass = {astro-ph.SR},
       adsurl = {https://ui.adsabs.harvard.edu/abs/2021ApJ...920....4M}
}

@ARTICLE{nordon2008,
       author = {{Nordon}, R. and {Behar}, E.},
        title = "{Abundance variations and first ionization potential trends during large stellar flares}",
      journal = {\aap},
         year = 2008,
        month = may,
       volume = {482},
       number = {2},
        pages = {639-651},
          doi = {10.1051/0004-6361:20078848},
archivePrefix = {arXiv},
       eprint = {0712.0482},
 primaryClass = {astro-ph},
       adsurl = {https://ui.adsabs.harvard.edu/abs/2008A&A...482..639N}
}

@ARTICLE{phillips2012,
       author = {{Phillips}, K.~J.~H. and {Dennis}, B.~R.},
        title = "{The Solar Flare Iron Abundance}",
      journal = {\apj},
         year = 2012,
        month = mar,
       volume = {748},
       number = {1},
          eid = {52},
        pages = {52},
          doi = {10.1088/0004-637X/748/1/52},
archivePrefix = {arXiv},
       eprint = {1201.2875},
 primaryClass = {astro-ph.SR},
       adsurl = {https://ui.adsabs.harvard.edu/abs/2012ApJ...748...52P}
}

@ARTICLE{polito2018,
       author = {{Polito}, Vanessa and {Galan}, Giselle and {Reeves}, Katharine K. and {Musset}, Sophie},
        title = "{Possible Signatures of a Termination Shock in the 2014 March 29 X-class Flare Observed by IRIS}",
      journal = {\apj},
         year = 2018,
        month = oct,
       volume = {865},
       number = {2},
          eid = {161},
        pages = {161},
          doi = {10.3847/1538-4357/aadada},
       adsurl = {https://ui.adsabs.harvard.edu/abs/2018ApJ...865..161P}
}

@ARTICLE{reep2020,
       author = {{Reep}, Jeffrey W. and {Antolin}, Patrick and {Bradshaw}, Stephen J.},
        title = "{Electron Beams Cannot Directly Produce Coronal Rain}",
      journal = {\apj},
         year = 2020,
        month = feb,
       volume = {890},
       number = {2},
          eid = {100},
        pages = {100},
          doi = {10.3847/1538-4357/ab6bdc},
archivePrefix = {arXiv},
       eprint = {2002.07669},
 primaryClass = {astro-ph.SR},
       adsurl = {https://ui.adsabs.harvard.edu/abs/2020ApJ...890..100R}
}

@ARTICLE{reep2024,
       author = {{Reep}, Jeffrey W. and {Unverferth}, John and {Barnes}, Will T. and {Chhabra}, Sherry},
        title = "{Modeling Time-variable Elemental Abundances in Coronal Loop Simulations}",
      journal = {\apjl},
         year = 2024,
        month = aug,
       volume = {970},
       number = {2},
          eid = {L41},
        pages = {L41},
          doi = {10.3847/2041-8213/ad64c3},
archivePrefix = {arXiv},
       eprint = {2406.04473},
 primaryClass = {astro-ph.SR},
       adsurl = {https://ui.adsabs.harvard.edu/abs/2024ApJ...970L..41R}
}

@article{reep2026,
    author = {Reep, Jeffrey and Benavitz, Luke Fushimi and To, Andy and Brooks, David H. and Laming, J. Martin and Antolin, Patrick and Long, David and Baker, Deborah},
    title = {Radiative hydrodynamic simulations of first ionization potential fractionation in solar flares},
    journal = {Philosophical Transactions of the Royal Society A: Mathematical, Physical and Engineering Sciences},
    volume = {384},
    number = {2323},
    pages = {20250189},
    year = {2026},
    month = {07},
    issn = {1364-503X},
    doi = {10.1098/rsta.2025.0189},
    url = {https://doi.org/10.1098/rsta.2025.0189},
    eprint = {https://royalsocietypublishing.org/rsta/article-pdf/doi/10.1098/rsta.2025.0189/6150015/rsta.2025.0189.pdf},
}

@ARTICLE{reeves2015,
       author = {{Reeves}, Katharine K. and {McCauley}, Patrick I. and {Tian}, Hui},
        title = "{Direct Observations of Magnetic Reconnection Outflow and CME Triggering in a Small Erupting Solar Prominence}",
      journal = {\apj},
         year = 2015,
        month = jul,
       volume = {807},
       number = {1},
          eid = {7},
        pages = {7},
          doi = {10.1088/0004-637X/807/1/7},
archivePrefix = {arXiv},
       eprint = {1505.07307},
 primaryClass = {astro-ph.SR},
       adsurl = {https://ui.adsabs.harvard.edu/abs/2015ApJ...807....7R}
}

@ARTICLE{savage2012,
       author = {{Savage}, Sabrina L. and {McKenzie}, David E. and {Reeves}, Katharine K.},
        title = "{Re-interpretation of Supra-arcade Downflows in Solar Flares}",
      journal = {\apjl},
         year = 2012,
        month = mar,
       volume = {747},
       number = {2},
          eid = {L40},
        pages = {L40},
          doi = {10.1088/2041-8205/747/2/L40},
archivePrefix = {arXiv},
       eprint = {1112.3088},
 primaryClass = {astro-ph.SR},
       adsurl = {https://ui.adsabs.harvard.edu/abs/2012ApJ...747L..40S}
}

@ARTICLE{sheeley2004,
       author = {{Sheeley}, Jr., N.~R. and {Warren}, H.~P. and {Wang}, Y.-M.},
        title = "{The Origin of Postflare Loops}",
      journal = {\apj},
         year = 2004,
        month = dec,
       volume = {616},
       number = {2},
        pages = {1224-1231},
          doi = {10.1086/425126},
       adsurl = {https://ui.adsabs.harvard.edu/abs/2004ApJ...616.1224S}
}

@INPROCEEDINGS{shimizu2021,
       author = {{Shimizu}, Toshifumi and {Imada}, Shinsuke and {Kawate}, Tomoko and {Suematsu}, Yoshinori and {Hara}, Hirohisa and {Tsuzuki}, Toshihiro and {Katsukawa}, Yukio and {Kubo}, Masahito and {Ishikawa}, Ryoko and {Watanabe}, Tetsuya and {Toriumi}, Shin and {Ichimoto}, Kiyoshi and {Nagata}, Shin'ichi and {Hasegawa}, Takahiro and {Yokoyama}, Takaaki and {Watanabe}, Kyoko and {Tsuno}, Katsuhiko and {Korendyke}, Clarence M. and {Warren}, Harry and {De Pontieu}, Bart and {Boerner}, Paul and {Solanki}, Sami K. and {Teriaca}, Luca and {Schuehle}, Udo and {Matthews}, Sarah and {Long}, David and {Thomas}, William and {Hancock}, Barry and {Reid}, Hamish and {Fludra}, Andrzej and {Auch{\`e}re}, Frederic and {Andretta}, Vincenzo and {Naletto}, Giampiero and {Poletto}, Luca and {Harra}, Louise},
        title = "{The Solar-C (EUVST) mission: the latest status}",
    booktitle = {Society of Photo-Optical Instrumentation Engineers (SPIE) Conference Series},
         year = 2021,
       editor = {{den Herder}, Jan-Willem A. and {Nikzad}, Shouleh and {Nakazawa}, Kazuhiro},
       series = {Society of Photo-Optical Instrumentation Engineers (SPIE) Conference Series},
       volume = {11444},
        month = jan,
          eid = {114440N},
        pages = {114440N},
          doi = {10.1117/12.2560887},
       adsurl = {https://ui.adsabs.harvard.edu/abs/2021SPIE11444E..0NS}
}

@ARTICLE{suzuki2025,
       author = {{Suzuki}, Hiromasa and {Kurashima}, Jun and {Katsuda}, Satoru and {Mori}, Koji and {Inoue}, Shun and {Ishi}, Daiki and {Churazov}, Eugene M. and {Khabibullin}, Ildar and {Sunyaev}, Rashid A. and {Mizuno}, Tsunefumi and {Kilbourne}, Caroline and {Ezoe}, Yuichiro and {Nakajima}, Hiroshi and {Sato}, Kosuke and {Miller}, Eric and {Matsushita}, Kyoko},
        title = "{XRISM observations of solar flare X-ray emission reflected in the Earth's atmosphere}",
      journal = {\pasj},
         year = 2025,
        month = dec,
       volume = {77},
       number = {6},
        pages = {1301-1312},
          doi = {10.1093/pasj/psaf108},
archivePrefix = {arXiv},
       eprint = {2509.05029},
 primaryClass = {astro-ph.SR},
       adsurl = {https://ui.adsabs.harvard.edu/abs/2025PASJ...77.1301S}
}

@ARTICLE{to2021,
       author = {{To}, Andy S.~H. and {Long}, David M. and {Baker}, Deborah and {Brooks}, David H. and {van Driel-Gesztelyi}, Lidia and {Laming}, J. Martin and {Valori}, Gherardo},
        title = "{The Evolution of Plasma Composition during a Solar Flare}",
      journal = {\apj},
         year = 2021,
        month = apr,
       volume = {911},
       number = {2},
          eid = {86},
        pages = {86},
          doi = {10.3847/1538-4357/abe85a},
archivePrefix = {arXiv},
       eprint = {2102.09985},
 primaryClass = {astro-ph.SR},
       adsurl = {https://ui.adsabs.harvard.edu/abs/2021ApJ...911...86T}
}

@ARTICLE{to2024,
       author = {{To}, Andy S.~H. and {Brooks}, David H. and {Imada}, Shinsuke and {French}, Ryan J. and {van Driel-Gesztelyi}, Lidia and {Baker}, Deborah and {Long}, David M. and {Ashfield}, IV, William and {Hayes}, Laura A.},
        title = "{Spatially resolved plasma composition evolution in a solar flare ─ The effect of reconnection outflow}",
      journal = {\aap},
         year = 2024,
        month = nov,
       volume = {691},
          eid = {A95},
        pages = {A95},
          doi = {10.1051/0004-6361/202449246},
archivePrefix = {arXiv},
       eprint = {2409.18188},
 primaryClass = {astro-ph.SR},
       adsurl = {https://ui.adsabs.harvard.edu/abs/2024A&A...691A..95T}
}

@article{to2026,
    author = {To, Andy S. H. and Laming, J. Martin and Reep, Jeffrey and Finley, Adam J.},
    title = {Chromospheric dynamics and turbulence regulate the solar FIP effect},
    journal = {Philosophical Transactions of the Royal Society A: Mathematical, Physical and Engineering Sciences},
    volume = {384},
    number = {2323},
    pages = {20250374},
    year = {2026},
    month = {07},
    issn = {1364-503X},
    doi = {10.1098/rsta.2025.0374},
    url = {https://doi.org/10.1098/rsta.2025.0374},
    eprint = {https://royalsocietypublishing.org/rsta/article-pdf/doi/10.1098/rsta.2025.0374/6149889/rsta20250374.pdf},
}

@ARTICLE{warren2011,
       author = {{Warren}, Harry P. and {O'Brien}, Casey M. and {Sheeley}, Jr., Neil R.},
        title = "{Observations of Reconnecting Flare Loops with the Atmospheric Imaging Assembly}",
      journal = {\apj},
         year = 2011,
        month = dec,
       volume = {742},
       number = {2},
          eid = {92},
        pages = {92},
          doi = {10.1088/0004-637X/742/2/92},
archivePrefix = {arXiv},
       eprint = {1109.2474},
 primaryClass = {astro-ph.SR},
       adsurl = {https://ui.adsabs.harvard.edu/abs/2011ApJ...742...92W}
}

@ARTICLE{warren2014,
       author = {{Warren}, Harry P.},
        title = "{Measurements of Absolute Abundances in Solar Flares}",
      journal = {\apjl},
         year = 2014,
        month = may,
       volume = {786},
       number = {1},
          eid = {L2},
        pages = {L2},
          doi = {10.1088/2041-8205/786/1/L2},
archivePrefix = {arXiv},
       eprint = {1310.4765},
 primaryClass = {astro-ph.SR},
       adsurl = {https://ui.adsabs.harvard.edu/abs/2014ApJ...786L...2W}
}

@ARTICLE{warren2018,
       author = {{Warren}, Harry P. and {Brooks}, David H. and {Ugarte-Urra}, Ignacio and {Reep}, Jeffrey W. and {Crump}, Nicholas A. and {Doschek}, George A.},
        title = "{Spectroscopic Observations of Current Sheet Formation and Evolution}",
      journal = {\apj},
         year = 2018,
        month = feb,
       volume = {854},
       number = {2},
          eid = {122},
        pages = {122},
          doi = {10.3847/1538-4357/aaa9b8},
archivePrefix = {arXiv},
       eprint = {1711.10826},
 primaryClass = {astro-ph.SR},
       adsurl = {https://ui.adsabs.harvard.edu/abs/2018ApJ...854..122W}
}

@ARTICLE{xie2022,
       author = {{Xie}, Xiaoyan and {Reeves}, Katharine K. and {Shen}, Chengcai and {Ingram}, Joshua D.},
        title = "{Statistical Study of the Kinetic Features of Supra-arcade Downflows Detected from Multiple Solar Flares}",
      journal = {\apj},
         year = 2022,
        month = jul,
       volume = {933},
       number = {1},
          eid = {15},
        pages = {15},
          doi = {10.3847/1538-4357/ac695d},
       adsurl = {https://ui.adsabs.harvard.edu/abs/2022ApJ...933...15X}
}

@ARTICLE{xie2023,
       author = {{Xie}, Xiaoyan and {Reeves}, Katharine K.},
        title = "{Heating Effects of Supra-arcade Downflows on Plasma above Solar Flare Arcades}",
      journal = {\apj},
         year = 2023,
        month = jan,
       volume = {942},
       number = {1},
          eid = {28},
        pages = {28},
          doi = {10.3847/1538-4357/ac9f47},
       adsurl = {https://ui.adsabs.harvard.edu/abs/2023ApJ...942...28X}
}

@ARTICLE{yu2020,
       author = {{Yu}, Sijie and {Chen}, Bin and {Reeves}, Katharine K. and {Gary}, Dale E. and {Musset}, Sophie and {Fleishman}, Gregory D. and {Nita}, Gelu M. and {Glesener}, Lindsay},
        title = "{Magnetic Reconnection during the Post-impulsive Phase of a Long-duration Solar Flare: Bidirectional Outflows as a Cause of Microwave and X-Ray Bursts}",
      journal = {\apj},
         year = 2020,
        month = sep,
       volume = {900},
       number = {1},
          eid = {17},
        pages = {17},
          doi = {10.3847/1538-4357/aba8a6},
archivePrefix = {arXiv},
       eprint = {2007.10443},
 primaryClass = {astro-ph.SR},
       adsurl = {https://ui.adsabs.harvard.edu/abs/2020ApJ...900...17Y}
}

@ARTICLE{zhu2016,
       author = {{Zhu}, Chunming and {Liu}, Rui and {Alexander}, David and {McAteer}, R.~T. James},
        title = "{Observation of the Evolution of a Current Sheet in a Solar Flare}",
      journal = {\apjl},
         year = 2016,
        month = apr,
       volume = {821},
       number = {2},
          eid = {L29},
        pages = {L29},
          doi = {10.3847/2041-8205/821/2/L29},
archivePrefix = {arXiv},
       eprint = {1603.07062},
 primaryClass = {astro-ph.SR},
       adsurl = {https://ui.adsabs.harvard.edu/abs/2016ApJ...821L..29Z}
}
\bibliographystyle{aasjournal}

\end{document}